\documentclass[structabstract]{aa}  
\usepackage{graphicx}
\usepackage{txfonts}
\usepackage{longtable} 
\begin{document}
   \title{`Ultimate' information content in solar and stellar spectra:}

   \subtitle{Photospheric line asymmetries and wavelength shifts}

   \author{Dainis Dravins}

   \institute{Lund Observatory, Box 43, SE-22100 Lund\\
              \email{dainis@astro.lu.se}}
        
   \date{Received July 1, 2008; accepted September 9, 2008}

% \abstract{}{}{}{}{} 
% 5 {} token are mandatory
 
  \abstract
   {Spectral-line asymmetries (displayed as bisectors) and wavelength shifts are signatures of the hydrodynamics in solar and stellar atmospheres.  Theory may precisely predict idealized lines, but accuracies in real observed spectra are limited by blends, few suitable lines, imprecise laboratory wavelengths, and instrumental imperfections.}
   {We extract bisectors and shifts until the `ultimate' accuracy limits in highest-quality solar and stellar spectra, so as to understand the various limits set by (i) stellar physics (number of relevant spectral lines, effects of blends, rotational line broadening), by (ii) observational techniques (spectral resolution, photometric noise), and by (iii) limitations in laboratory data.}
   {Several spectral atlases of the Sun and bright solar-type stars were examined for those thousands of `unblended' lines with the most accurate laboratory wavelengths, yielding bisectors and shifts as averages over groups of similar lines.  Representative data were obtained as averages over groups of similar lines, thus minimizing the effects of photometric noise and of random blends.}
   {For the solar-disk center and integrated sunlight, the bisector shapes and shifts were extracted for previously little-studied species (\ion{Fe}{ii}, \ion{Ti}{i}, \ion{Ti}{ii}, \ion{Cr}{ii}, \ion{Ca}{i}, \ion{C}{i}), using recently determined and very accurate laboratory wavelengths.  In Procyon and other F-type stars, a sharp blueward bend in the bisector near the spectral continuum is confirmed, revealing line saturation and damping wings in upward-moving photospheric granules. Accuracy limits are discussed: `astrophysical' noise due to few measurable lines, finite instrumental resolution, superposed telluric absorption, inaccurate laboratory wavelengths, and calibration noise in spectrometers, together limiting absolute lineshift studies to $\approx$~50-100~m~s$^{-1}$.}
   {Spectroscopy with resolutions $\lambda$/$\Delta\lambda\approx$~300,000 and accurate wavelength calibration will enable bisector studies for many stars.  Circumventing remaining limits of astrophysical noise in line-blends and rotationally smeared profiles may ultimately require spectroscopy across spatially resolved stellar disks, utilizing optical interferometers and extremely large telescopes of the future.}


\titlerunning{Information content in solar and stellar spectra}
\authorrunning{D.Dravins}
\maketitle

\section{Preamble and context}

Numeric simulations of three-dimensional and time-dependent hydrodynamics have become established as a realistic description for the convective photospheres of the Sun and solar-type stars.  Such models reproduce the observed fine structure (granulation) across the solar surface, its time evolution, as well as its interaction with magnetic fields.  Using the output from such simulations as temporally and spatially varying model atmospheres, synthetic spectral line profiles are computed as temporal and spatial averages, accurately reproducing spectral-line profiles without invoking any free fitting parameters such as `mixing-length' or `turbulence', used extensively in the past.  For reviews of such analyses, see Dravins (1982) and Asplund (2005). 

These simulations have also brought about a certain change in paradigm.  Contrary to previous practice, it has been realized that it is not possible, not even in principle, to infer detailed properties of some stellar atmosphere from an analysis of the observed line parameters alone, no matter how precisely the spectrum would be measured.  Any photospheric line is built by a great many different contributions from a wide variety of temporally variable lateral inhomogeneities across the stellar surface, whose statistical averaging over time and space produce the line shapes and shifts that can be observed in integrated starlight.  While it is possible to compute the resulting line profiles and bisectors from hydrodynamic models, the opposite is not feasible because the three-dimensional atmospheric structure cannot be uniquely deduced from observed lineshapes alone.

Reliable prediction of synthetic line profiles is essential not only for studying stellar atmospheres per se, but also when using stars as probes for chemical evolution in the Galaxy or as test bodies for binary motion with low-mass companions.  For example, determining the isotopic abundance ratio $^{6}$Li/$^{7}$Li requires analyzing slight absorption-line asymmetries, disturbingly similar to the asymmetries produced by stellar surface dynamics (Asplund et al.\ 2006; Cayrel et al.\ 2007; Nissen 2000).  Understanding the surfaces of stars such as the $\alpha$~Cen components or of Procyon is required both for analyzing stellar structure (Allende Prieto et al.\ 2002a; Frutiger et al.\ 2005; Robinson et al.\ 2005), and for interpreting oscillations and stellar limb darkening (Aufdenberg et al.\ 2005; Bigot et al.\ 2006).  The latter is an important parameter for interferometric observations.  Successively improved laboratory data are also enabling precise line synthesis for species that have not been much studied in the past, including those with complex atomic hyperfine structure such as the iron-group element manganese (Bergemann \& Gehren 2007) or the rare-earth element praseodymium (Ivarsson et al.\ 2003).

While observations resolve the overall structure of solar granulation (enabling comparison with models), recent simulations have improved the spatial resolution all the way to a few kilometers, revealing significant phenomena even on these astrophysically miniscule scales, such as magnetic dynamo processes or acoustic shocks (Muthsam et al.\ 2007; Sch\"{u}ssler \& V\"{o}gler 2008; V\"{o}gler \& Sch\"{u}ssler 2007).  Corresponding observational resolutions remain unrealistic for the foreseeable future, but information on even such small-scale phenomena could in principle be extracted from spatially-averaged line asymmetries and wavelength shifts.  Signatures from structures below the spatial resolution limit may be identified by comparing simulated spectral line parameters from models including specific phenomena (shock waves, etc.) to models without them.

Although hydrodynamic simulations should not have freely adjustable parameters, their computational complexity implies that they must contain various physical, mathematical, and numerical approximations to be manageable.  Predicted spectral-line parameters are sensitive to the details of model simulations (e.g., Asplund et al.\ 2000a), but unless there is some independent verification of the validity of the model approximations, the significance of their predictions may remain in doubt.

Arguably, the most important aspect in testing any model prediction is the ability to falsify it.  For hydrodynamic atmospheres (other than the spatially resolved Sun), prime parameters capable of falsifying models are line asymmetries and wavelength shifts of photospheric absorption lines.  These have no equivalent in classical modeling (where parameters such as `turbulence' cause only a symmetric line broadening about the central wavelength), but in more realistic models they originate from the statistical bias of a greater number of [spectrally blueshifted] photons from hot and rising surface elements (granules) than from the cooler and sinking ones.  Since the velocity fields vary with depth, and the line strength depends on local temperature and pressure, the resulting line asymmetries and shifts depend on spectral-line parameters, such as the oscillator strength, the excitation potential, ionization level, and the wavelength region.  Obviously, simulations that reproduce observed spectral-line signatures for different classes of lines are more trustworthy applied to other purposes.

Following years of progress, one is now approaching a somewhat paradoxical situation.  On one hand, hydrodynamic models can predict detailed properties for a hypothetical spectral line of any given atomic species in some particular ionization stage, with a chosen excitation potential, in some specified wavelength region, etc.  On the other hand, the crucial confrontation with observations may no longer be possible, because corresponding stellar lines of the desired strength, ionization level, and wavelength region may simply not exist in nature, or are unobservable in practice!  In real stellar spectra, lines are frequently blended by stellar rotation or by overlapping telluric lines from the terrestrial atmosphere or else smeared by inadequate instrumental resolution and insufficient spectral fidelity, in practice precluding any detailed study.

Understanding the accuracy limits for absolute wavelength shifts is required not only in studying convection signatures but also for possible long-term changes in stellar, interstellar, or intergalactic absorption features (perhaps caused by stellar activity cycles or by the changing expansion of the Universe), such as envisioned in programs with instruments for the future extremely large telescopes (Pasquini et al.\ 2005; D'Odorico et al.\ 2007).  Especially if measurements from different spectrometers at different epochs are to be compared, observational procedures must be optimized with an understanding of what sets the limits.  

Exploring the accuracy limits for solar and stellar wavelengths appears to require some even more accurate external reference.  Here, the patterns of astrophysical wavelength shifts in solar lines themselves will be used, as referring to laboratory wavelengths.  This still is not without problems since the exact solar shifts have not been reliably calculated for all various lines, and the inaccuracies in laboratory wavelengths are often not understood.  Nevertheless, it is possible to search for [in]consistencies in the shifts of astrophysically similar lines, measure differences between recordings from different instruments, and then try to trace their origins.

In this work, accuracies of wavelength shifts are studied, as referring to absolute wavelength measures.  The somewhat related problem of relative precision (i.e., reproducibility) is considered in the context of, e.g., exoplanet searches through radial-velocity techniques and is not the issue here.  The internal precision can be much higher than the external accuracy, and even the most precise of today's radial-velocity instruments have internal zero-point differences (between, e.g., the wavelength scales in their different spectral orders) several orders of magnitude worse than their internal stability.

\section{Solar and stellar spectra}

It is not known what resolution and accuracy parameters stellar spectra ideally should have for their information content to be `fully' exploited.  However, existing solar spectral atlases with resolutions not less than R~=~$\lambda$/$\Delta\lambda\approx$~300,000, and signal-to-noise ratios on the order of 3,000 have been very productive for numerous studies.  Ideally, we would also desire stellar atlases with such spectral fidelity, with matching accuracies in wavelength calibrations, for spectra with minimal rotational broadening (although narrow features remain also in the spectra of rapidly rotating stars; Dravins et al.\ 1990c).  The reality, of course, is that stellar data with this level of quality do not yet exist.

Following some examination of various spectra also in the infrared, the wavelength range to be studied was constrained to only the visual.  A limit towards shorter wavelengths is set by increased line blending in the (ultra)violet, which leaves virtually no unblended lines, while a longward limit arises from increased telluric absorption that blends many among the relatively few (and mostly weak) stellar lines in the near infrared.

\subsection{Solar spectral atlases}

Spectra of both solar disk center and such of integrated sunlight (i.e., the Sun seen as a star) were analyzed.  Since line shapes and shifts change across the disk towards the limb, and lines are broadened by solar rotation, those spectra are noticeably different.  

For the disk center, two atlases were analyzed. The first is the `Jungfraujoch Atlas' by Delbouille et al.\ (1989), covering 300-1000 nm and recorded with a double-pass grating spectrometer with resolving power R~$\approx$~700,000 and small-scale photometric random noise on the order of 1/2000 of the signal.  However, possible systematics in the normalization of more extended spectral regions could be more awkward to quantify (for some example, see Dravins 1994).

The second is part of the `Spectral Atlas of Solar Absolute Disk-Averaged and Disk-Center Intensity from 3290 to 12510 \AA', recorded with the Fourier Transform Spectrometer (FTS) at Kitt Peak National Observatory, going back to an unpublished version by Brault and Neckel.  Here a more processed version is used, made available by the Hamburg Observatory (Neckel 1999).  We refer to this as the `Hamburg photosphere'.  Spectral resolving power here is about 350,000, and a typical ratio of signal to small-scale random noise about 3,000.

For integrated sunlight, the `Solar flux atlas from 296 to 1300 nm' by Kurucz et al.\ (1984) is used, which we refer to as the `Kitt Peak Atlas'.  Its resolving power ranges between about 350,000 and 500,000, with the peak ratio of signal to random photometric noise again on the order of 3,000.

\subsection{Stellar spectra}

No stellar spectra have yet been recorded with the fidelity of solar atlases.  For a few bright stars (Arcturus, Procyon, Sirius, etc.), atlases have been produced in formats reminiscent of the solar ones, but their resolution and noise levels are inferior by factors of several.  Also, these represent only a few spectral types across the Hertzsprung-Russell diagram.

\subsubsection{ESO UVES Paranal project}

For a reasonably complete and homogeneous sample of different spectral types, recorded with a quality that is representative of current high-fidelity stellar spectroscopy, we used `A Library of High-Resolution Spectra of Stars across the Hertzsprung-Russell Diagram' from the UVES Paranal Project (Bagnulo et al.\ 2003).  Numerous stars were observed with the UVES spectrometer (Dekker et al.\ 2000) at the ESO Very Large Telescope {\it{Kueyen}} unit.  The interval 300--1000 nm was covered at a resolution of R~$\approx$~80,000, optimizing the data analysis in a non-standard manner for reaching the highest reasonably possible signal-to-noise ratios of $\approx$~300--500.  While the usual UVES data reduction `pipeline' is optimized for S/N ratios on the order of 100, a special reduction package was developed for this program of very bright and well-exposed stars to permit these much higher values.

Although somewhat better stellar data may exist in a few cases, the UVES Paranal Project appears to offer the currently best coverage of high-quality stellar spectra surrounding the Sun in the HR-diagram.  The resolution of the UVES spectrometer itself can be pushed to somewhat higher values (at least in its red arm), but the differences would only be slight (compared to the more wishful step to reach solar-atlas standards).  Thus, these data are not likely to be substantially surpassed until seriously high-resolution instruments become available, such as PEPSI on the Large Binocular Telescope (Andersen et al.\ 2005).

\subsubsection{Procyon and Arcturus}

In a few cases, stellar spectra with superior resolution are available.  One case is Procyon (F5 IV-V), where R~$\approx$~200,000 spectra were recorded in certain spectral segments at the McDonald Observatory (Allende Prieto 1999) and where some individual lines had already earlier been observed with similar resolutions with an ESO double-pass scanner (Dravins 1987b).  The older (photographically recorded) atlas of Procyon (Griffin \& Griffin 1979) does not quite match the photometric noise (ca.\ 1-2\%), but its extensive wavelength coverage and careful preparation still make it a useful reference (with resolution R~$\approx$~160,000 better than UVES Paranal) and a source for double-checking specific line profiles.

The probably highest fidelity stellar spectrum currently available is that of the K-giant Arcturus (Hinkle et al.\ 2000).  However, this will not be discussed further here because extensive line blending in the rich spectra of such cooler K-type stars appears to require somewhat different types of analyses.

\subsection{Spectral line selection}

From such spectral `atlases', the next step is to select individual lines for the statistical analyses.  Line bisectors can be obtained from photometrically precise intensity profiles, but their wavelength placement requires accurate laboratory wavelengths, constraining the number of candidate lines.

\subsubsection{Candidate solar and stellar lines}

Any more complete (and reliable) identification of the many thousands of lines potentially identifiable in each stellar spectrum is a task beyond the scope of this paper, which is why other approaches were chosen.  For selecting samples of lines, we start from solar spectra, for which line identifications are available from earlier wavelength tables and atlases and, for relatively unblended lines, in particular from Allende Prieto \& Garc{\'\i}a L\'{o}pez (1998), who list a total of 4947 in the range 394--796 nm.  This limits stellar samples to subsets of those lines that also exist in the Sun, but should be justified since we only study stars not very far from the Sun in the HR-diagram.  Also, comparing the same line in different stars should be more straightforward than comparing different lines in different stars.  A similar view was taken by Gray (2005) in making a survey of the bisector shapes for one particular \ion{Fe}{i} line in numerous G- and K-type stars.

The by far largest subgroup of lines in this listing is made up of \ion{Fe}{i} with 2072 solar ones, followed by \ion{Cr}{i} with 535 lines, \ion{Ti}{i} (457), \ion{Ni}{i} (393), \ion{V}{i} (189), \ion{Co}{i} (163), \ion{Fe}{ii} (139), \ion{Ti}{ii} (104), \ion{Mn}{i} (100), \ion{Si}{i} (91), \ion{Ca}{i} (76), \ion{Ce}{ii} (62), \ion{Cr}{ii} (64), and \ion{Nd}{ii} (53).  Forty other species have fewer than 50 lines each.

\begin{table*}
\caption{Stars and their spectra that were selected for studying line asymmetries and wavelength shifts.}             
\label{table:1}      
\centering          
\begin{tabular}{c c c c}
\hline\hline       
Object & Type\footnote{1} & Spectral data & Notes \\ 
\hline                    
\object{68 Eri} &	F2 V &	UVES Paranal & $Vsini$ $\approx$~5 kms-1 \\
\object{$\theta$ Scl} & F5 V & UVES Paranal & $Vsini$ $\approx$~3 kms-1 \\
\object{Procyon}	& F5 IV-V & UVES Paranal	& R = 80,000 \\
Procyon & F5 IV-V &	Griffin \& Griffin & R = 160,000 \\
Procyon & F5 IV-V &	McDonald & R = 200,000 \\
HD 122563  & F8 IV &	UVES Paranal& [Fe/H] $\approx$~--2.7 \\  
\object{$\nu$ Phe} & F9 V &	UVES Paranal	\\
$\zeta$ Tuc & F9 V &	UVES Paranal	\\
HD 20807 & G1 V & UVES Paranal	\\
Solar disk center & G2 V &	Jungfraujoch & `Jungfraujoch Atlas' \\	
Solar disk center &	G2 V	& Kitt Peak/Hamburg &	`Hamburg photosphere' \\
Solar integrated flux	& G2 V	& Kitt Peak	& `Kitt Peak Atlas' \\
Moon & G2 V & UVES Paranal \\
HD 76932 & G2 V & UVES Paranal & [Fe/H] $\approx$~--1.0 \\
$\beta$ Hyi & G2 IV & UVES Paranal	\\
HD 59468 & G6 V & UVES Paranal \\
61 Vir & G7 V & UVES Paranal \\
HD 104304 & G8 IV &	UVES Paranal \\	
HD 24616	& G8 IV-V & UVES Paranal	\\
\hline                  
\end{tabular}
\begin{list}{}{}
\item[$^{\mathrm{1}}$] from SIMBAD (Centre de Donn\'{e}es Astronomiques de Strasbourg) 
\end{list}
\end{table*}

\subsubsection{Joint astronomical and laboratory data}

Since convective shifts (in at least dwarf stars) are expected to be no more than a few hundred m~s$^{-1}$, and relative shifts between groups of different lines only a fraction of that, the desired laboratory data contribution to the error budget should be less than some 100~m~s$^{-1}$.  Although extensive line lists exist in the literature, their accuracies generally are inadequate by about one decimal place.  The literature was searched for laboratory data and direct contacts made with some laboratory spectroscopy groups to gain access to unpublished measurements.  In some cases, line wavelengths had been measured, but their publication was awaiting a parallel (often time-consuming) solution for the atomic energy levels for all corresponding atomic transitions.  For our application, however, precise values for the latter are not required.

The following data sets with spectral lines jointly satisfying the criteria for both solar and laboratory spectra could then be identified: \ion{Ca}{i} -- 47 joint lines (laboratory data from Litz\'{e}n, private comm.); \ion{Co}{i} -- 153 (Pickering \& Thorne 1996, \& Pickering, private comm.); \ion{Cr}{i} -- 471 (Johansson, private comm.); \ion{Cr}{ii} -- 57 (Johansson, private comm.); \ion{Fe}{i} -- 1507 (Nave et al.\ 1994); \ion{Fe}{ii} -- 137 (Johansson, private comm.); \ion{Ni}{i} -- 369 (Litz\'{e}n et al.\ 1993); \ion{Si}{i} -- 1 (one line only; Engleman, private comm.); \ion{Ti}{i} -- 429 (Forsberg 1991); \ion{Ti}{ii} -- 73 (Zapadlik et al. 1995; Zapadlik 1996, \& Johansson, private comm.).  The total number is thus 3244.

The most prevalent species where more accurate laboratory data would be desirable are \ion{V}{i} with 189 solar lines, followed by \ion{Mn}{i} with 100.  However, similar to \ion{Co}{i}, these odd-Z iron-group elements come with the challenging aspect of a significant hyperfine structure in their lines.  Despite unclear levels of laboratory wavelength accuracy, a few more line groups were retained, e.g., lines from \ion{C}{i} where the small atomic mass causes much greater thermal line broadening than for heavier metals, thus possibly probing some different parts of parameter space. 

\subsubsection{Selecting stellar spectra}

Sample spectral regions in the UVES Paranal spectra of several tens of A-, F-, G-, and K-type stars were examined and bisectors computed for representative spectral lines to estimate what measurement precisions would be reachable for each star.  The conclusions, however, were somewhat disappointing in showing that -- although these spectra are excellent by normal stellar standards and the spectral resolution adequate for resolving at least the broader lines in at least the earlier-type stars -- by solar standards they are still very noisy, impeding several classes of worthy studies. 

For earlier-type (A- and early F) stars, the limit is set by the widths of the spectral lines: already intrinsically broad and often further broadened by rapid stellar rotation.  Multiple factors conspire against bisector studies: small intensity gradients in broad lines worsen the noise in measured bisector shapes, intrinsically broad lines already imply some astrophysical averaging of line asymmetries across the stellar surface, rotational broadening produces smearing with blending lines, the strength of absorption lines is intrinsically low, and the number of measurable lines is small.  For these reasons, no A-type star could be retained in the sample.

For later-type (G- and K) stars, different limits set in.  The very high density of absorption lines (especially at shorter wavelengths) implies that, even if the line cores might be reasonably unblended, almost all line wings are affected by blends that strongly distort the bisector shapes.  Furthermore, line crowding starts to make line identifications uncertain, possibly requiring quite detailed atmospheric model calculations for reliable classification.  For these reasons, no K-type star could be retained in the sample.  This study therefore targets F- and G-type stars with slow projected rotational velocities.  The retained datasets are listed in Table~1.

Besides `ordinary' stars, two metal-poor ones were also included with the suspicion that their bisectors might be different: HD76932 and HD122563.  For previous observations, see Allende Prieto et al.\ (1999), and for a theoretical discussion, see Shchukina et al.\ (2005).  However, their spectra are noisier than for the others and, in addition, bisector measurements are made awkward due to the intrinsic faintness of their absorption lines.

\section{Analyses of line profiles and bisectors}

With 3244 joint solar and laboratory lines and 19 solar/stellar spectra, tens of thousands of individual line profiles are candidates for analysis.  An experience from earlier solar work is that `blind' averaging of stellar line parameters must be avoided lest one contaminate the sample with too many outliers, which may easily hide physical trends.  However, the separation of outliers from representative data is a somewhat laborious task that is awkward to automate.  Decisions have to be made as to which line profiles and wavelengths are sufficiently undisturbed to become parts of statistical averages and which are too contaminated.

\subsection{Line-profile and bisector examination}

As a first step, individual line profiles of each species in each star were plotted (or, more precisely, a narrow wavelength region of the observed spectrum where that particular line was expected).  Several thousand line profiles were thus plotted, however not all lines in all stars.  Examples are in Fig.~1, where the same three \ion{Fe}{i} lines are seen in four different solar spectra: two disk-center versions, integrated sunlight observed with a solar spectrometer, and moonlight with a stellar instrument.  Already the modest solar rotational broadening of $\approx$~2~km~s$^{-1}$ (plus the effects of intrinsic center-to-limb changes) lead to quite visible differences between the spectra at solar disk center and that of integrated sunlight, illustrating the usefulness of slowly rotating stars.  The UVES spectrum of moonlight shows the degradation that affects even such high-quality stellar recordings, making clear that substantial enhancements will be required to obtain spectra of a fidelity that approaches the solar ones.

\begin{figure}
\centering
\includegraphics[width=9cm]{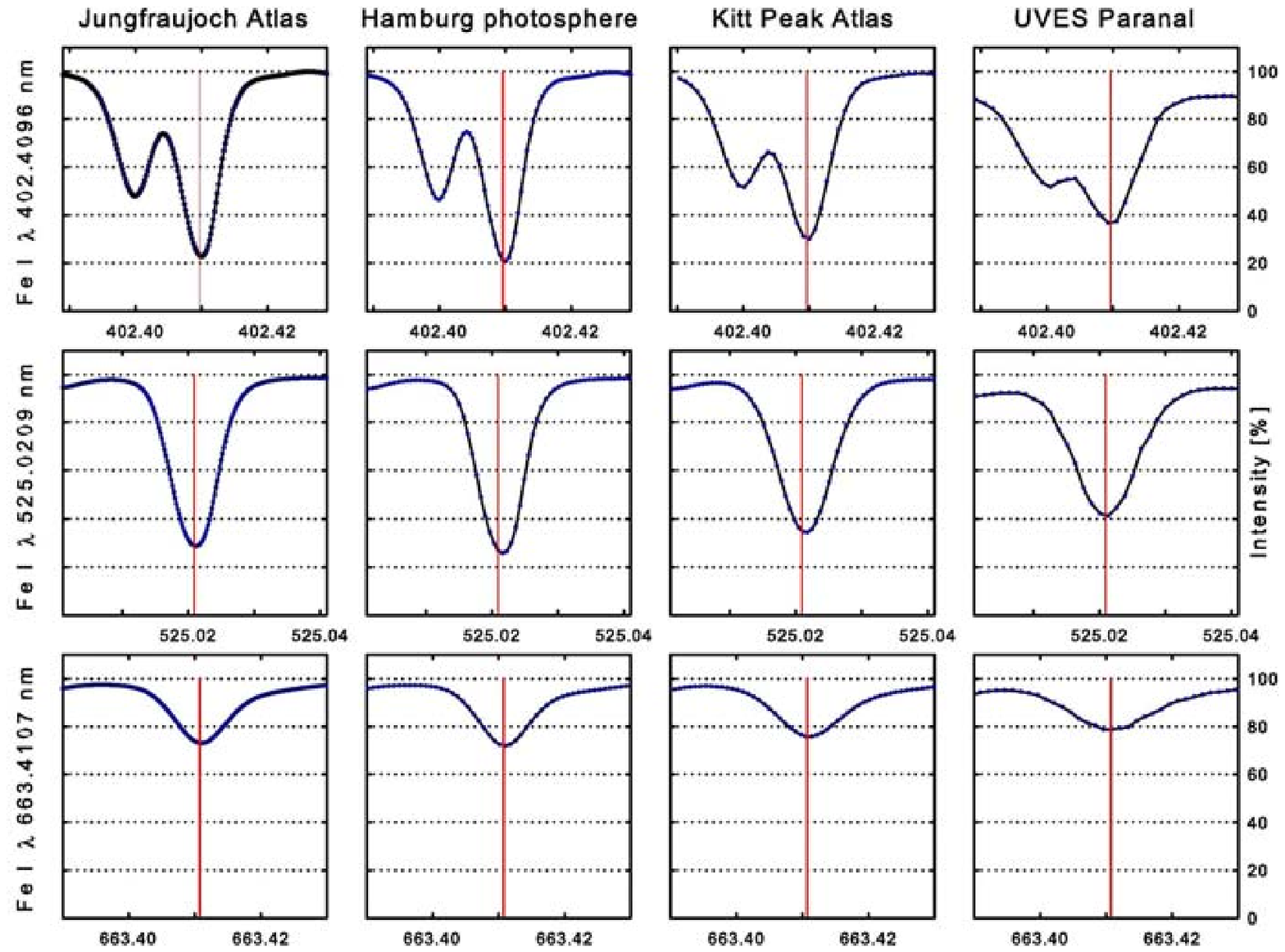}
\caption{Examples of individual line profiles in four different recordings of solar spectra.  From the top down, three representative \ion{Fe}{i} lines are seen: a deep but somewhat blended line in the violet, a clean line in the green, and a weaker line in the red.  Left to right: solar disk center (Jungfraujoch Atlas), solar disk center (Hamburg photosphere), integrated sunlight (Kitt Peak Atlas), moonlight (UVES Paranal).  Vertical lines mark laboratory wavelengths and thus the `naively' expected line positions.}
\label{Fig1}
\end{figure}

As the next step, these thousands of profiles were manually examined to verify whether the expected line was visible, and if so, which parts of it appeared undisturbed enough for the calculation of a bisector.  (Often,  the bisector can be well defined in the core of the line, but a blend in a line wing causes it to fly off closer to the continuum intensity level.)  The bisector was then computed for the selected part of each line and plotted, both individually and together with other bisectors of that particular species in that one star.

Representative bisectors are in Fig.~2, showing four \ion{Fe}{ii} lines in the stars \object{68~Eri}, \object{$\theta$~Scl}, and \object{$\nu$~Phe} of the sequential spectral classes F2~V, F5~V, and F9~V.  Comparing the same line between different stars, it is seen that bisectors often share common features, but that the bisector shapes differ strongly among different lines in the same star.  Thus, the `noise', i.e. the cause of bisector deviations from a representative mean, does not originate from photometric errors but instead is largely `astrophysical' in character, caused by blending lines and similar.

\begin{figure}
\centering
\includegraphics[width=8.8cm]{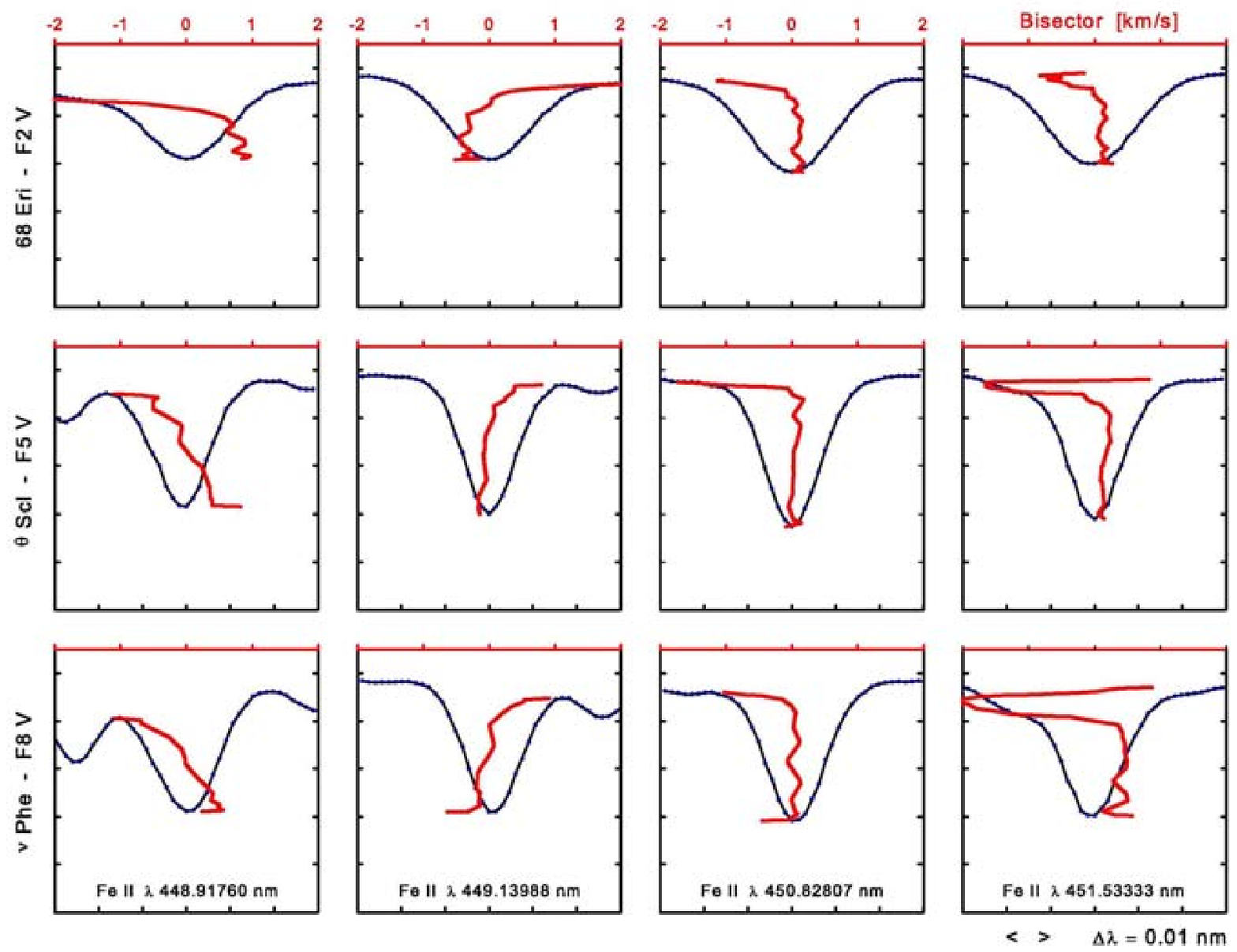}
\caption{Examples of individual bisectors, overplotted on the line profiles, for three representative \ion{Fe}{ii} lines in UVES Paranal spectra of different F-type stars: 68~Eri (F2~V), $\theta$~Scl (F5~V), and $\nu$~Phe (F9~V).  The bisector scale (top) is expanded a factor of 10 relative to the wavelength scale (shown as wavelength intervals $\Delta\lambda$; exact wavelength values depend on stellar radial motion.)}
\label{Fig2}
\end{figure}

It actually is this realization (also gained in earlier solar studies) that forms the basis of the present project.  The omnipresence of weak line-blends makes it impossible to extract reliable asymmetries or shifts from measurements of any one single line (or a group of only a few).  Even if a visual inspection suggests that some line appears `unblended' with an apparently smooth shape and bisector, the asymmetry may well be fortuitous, since other lines, appearing equally `unblended', may show different type of asymmetries.  Although studies of few individual bisectors thus may not be very conclusive, such should be practicable from larger groups of lines if one can somehow average away the noise originating from blends and inexact laboratory wavelengths.

Once a representative average bisector is determined, it allows some specific lines to be identified whose bisector shapes and/or shifts fall close to this average and that could be expected to also be representative of the intensity profiles.  A possible caveat remains in that the profile could still be misleading if the line is symmetrically blended in both its short- and long-wavelength wings.  This again calls for the analysis of several analogous lines (also because a bisector is a function only of line-depth, and lines of a given absorption depth can have different widths and shapes).

\subsection{Selection of line-groups}

Solar studies show how three-dimensional atmospheric structures leave different imprints on different classes of lines.  The most important variable is line-strength: asymmetries and shifts change with absorption-line depth (lines of different strength sample somewhat different atmospheric volumes).  Within groups of lines with the same strength, weaker dependences exist on the ionization level, the atomic excitation-potential, and the wavelength region (Dravins et al., 1981; 1986; Dravins \& Nordlund 1990ab; Gadun 1994; Asplund et al.\ 2000b).

However (except marginally for the Sun), observed spectra are not adequate for subdividing the sample with respect to all these parameters.  Even the dependence on linestrength is problematic since really weak lines are challenging to measure due to the shallow slopes of their wings.  A so-called `very good' signal-to-noise level of 400, say, (referring to the 100\% intensity of the continuum) for a weak line of 5\% absorption depth implies a noise level equal to 1:10 of that line's absorption half-depth, so completely inadequate for bisector studies.

For each star, line groups were formed separately for each atomic species and each ionization level.  We next present representative results for such different spectra in different stars.

\section{Solar line asymmetries and shifts}

Statistics of solar iron-line bisectors on an absolute wavelength scale have been examined in the past using data from spectral atlases, in particular for \ion{Fe}{i}, but also to a lesser extent for \ion{Fe}{ii} (Dravins et al.\ 1981, 1986; Asplund et al.\ 2000b).  Of course, there is extensive literature on solar line asymmetries (however, often excluding the wavelength shifts) for various magnetic and non-magnetic features, at different center-to-limb locations, and their changes on timescales from minutes to years.

\subsection{\ion{Fe}{ii} with new laboratory wavelengths}

Laboratory accuracies are more challenging for ionized species: light sources typically need higher temperatures or stronger currents (thus becoming more affected by pressure-shifts and electric-field effects), the emission in the source is more inhomogeneous (hence the light path inside the spectrometer could be slightly deviant), and past laboratory wavelengths for \ion{Fe}{ii} have not been quite up to the level of neutral iron.  However, ongoing collaborations in improving laboratory data for iron-group elements (in particular the ongoing FERRUM Project; Johansson 2002) have attained significantly more precise atomic energy levels, and thus accurate laboratory wavelengths for many more \ion{Fe}{ii} lines than before.

Figure 3 uses these new wavelengths to put \ion{Fe}{ii} bisectors on an absolute wavelength scale for both solar disk center and integrated sunlight.  The total number of lines jointly in the Sun and in the laboratory dataset was 137 to begin with, out of which 104 were accepted as clean enough for solar disk center and 93 for integrated sunlight.  Within each group, the average bisector is calculated as the arithmetic mean of the positions of those bisectors that contribute at each absorption depth, sampled in steps of 0.1\% continuum intensity.  For any bisector to be well-defined requires not only good spectral resolution and low photometric noise but also adequate intensity gradients in the line profile.  The gradients become small close to the continuum and near the linebottoms, where the bisectors must be terminated lest they fly off to unphysical values of many km~s$^{-1}$.  One data point in such starting deviations has been retained to indicate the observational limits, here seen as horizontal streaks at the bottoms of some individual bisectors. (Each curve is typically made up of some 500 points, and the inclusion of these few last points does not perceptibly affect the average.)  To facilitate comparisons with earlier work, the scale of the plots (i.e., axes ratio between spectral intensity and wavelength shift) is kept similar to the one in our earlier \ion{Fe}{i} studies cited above.

\begin{figure}
\centering
\includegraphics[width=6cm]{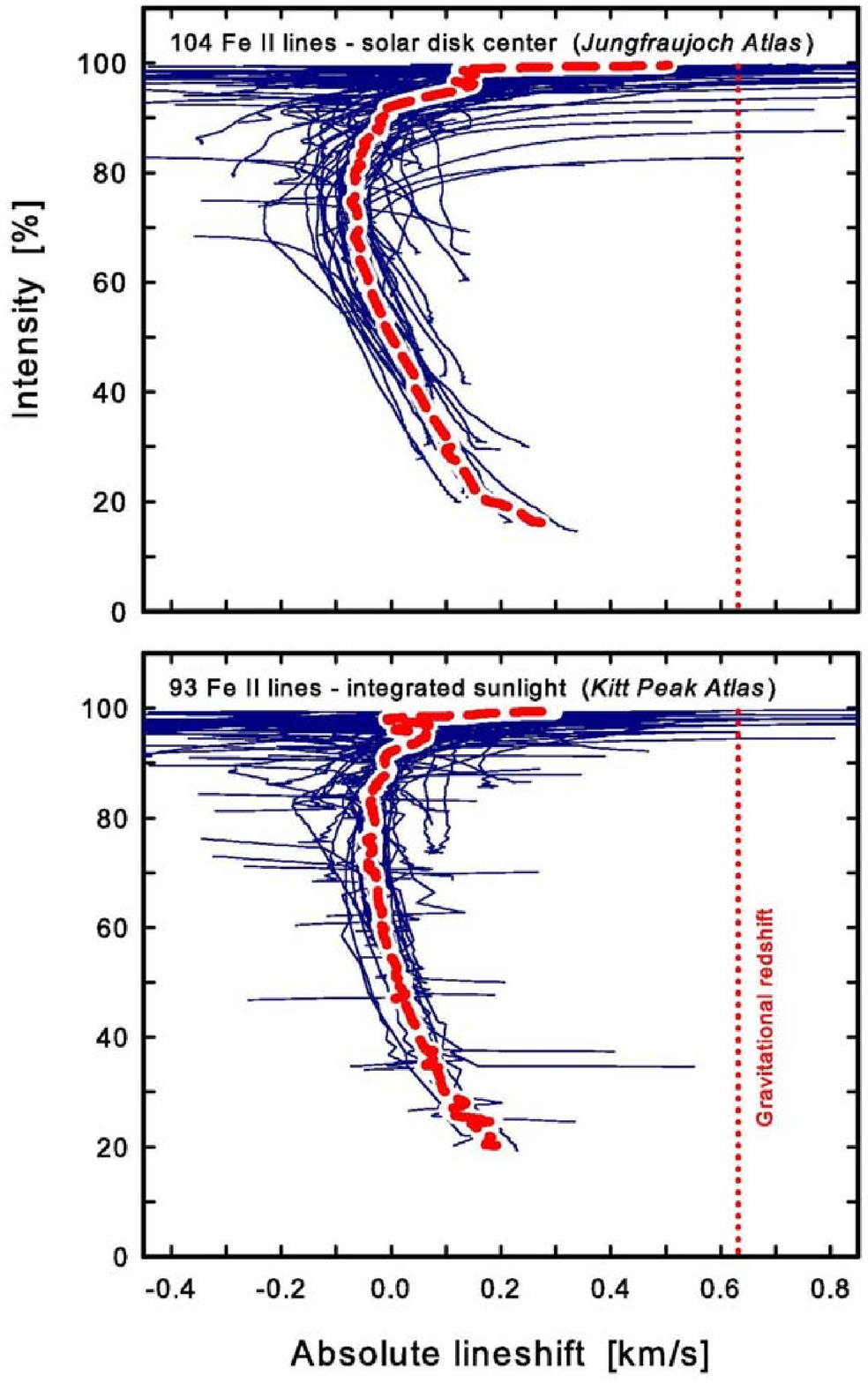}
\caption{\ion{Fe}{ii} bisectors in spectra of solar-disk center (Delbouille et al.\ 1989; top) and of integrated sunlight (Kurucz et al.\ 1984; bottom), placed on an absolute wavelength scale.  Each thin curve is the bisector of one spectral line; the thick dashed curve is their average.  The vertical scale denotes the percent of the intensity of the spectral continuum, while the dotted line marks the wavelength position expected in a classical (non-hydrodynamic) model atmosphere, given by the laboratory wavelength, displaced by the solar gravitational redshift for an Earth-bound observer of 633~m~s$^{-1}$.  Laboratory data are from the FERRUM Project (Johansson, private comm.)}
\label{Fig3}
\end{figure}

Figure 3 shows that the average bisector shapes are now well-defined, with a precision that can be judged from the spread around the mean.  The shapes do not differ much between the two datasets (except lines being somewhat deeper and more asymmetric at disk center), and also the amount of lineshift is comparable.  However, whether the wavelength shifts are affected by systematic offsets cannot be told from such data alone. 

Table 2 (in electronic version only) lists wavelengths for all \ion{Fe}{ii} lines examined, indicating whether the line was accepted as clean enough in each solar and stellar spectrum, and then plotted in Fig.~3 or in other figures below.  Laboratory wavelengths from the FERRUM Project are given there to five decimal places (in nm), one more than in traditional wavelength tables (the final decimal now corresponds to 6~m~s$^{-1}$ at $\lambda$~500 nm).  Note, however, that these wavelengths are taken from work in progress and -- since the exact atomic energy levels still may be subject to some refinement -- these values could eventually be superseded by slightly different ones.

\subsection{Solar \ion{Ti}{i} and \ion{Ti}{ii}}

For the solar disk-center, 310 \ion{Ti}{i} solar disk-center lines were retained from the original 429 joint solar and laboratory ones: Fig.~4. The line-bottom shift of the average bisector is now almost zero!  Although synthetic \ion{Ti}{i} lines from hydrodynamic models do not appear to have been published, it would be surprising if the shifts of their line-bottoms could be very much different from those of \ion{Fe}{i} (e.g., Dravins et al.\ 1981) or \ion{Fe}{ii} in Fig.~3.  While a physical origin cannot yet be completely excluded, one is led to suspect a zero-point offset in the laboratory wavelength scales.

The laboratory wavelengths originate from Forsberg (1991), but with one additional decimal extracted from the original data by Litz\'{e}n (private comm.), and are listed in the electronic Table 3.  Those wavelengths were measured with the Kitt Peak Fourier Transform Spectrometer (the very same instrument as was used for recording solar spectral atlases), using mainly argon lines for wavelength calibration.

\begin{figure}
\centering
\includegraphics[width=6cm]{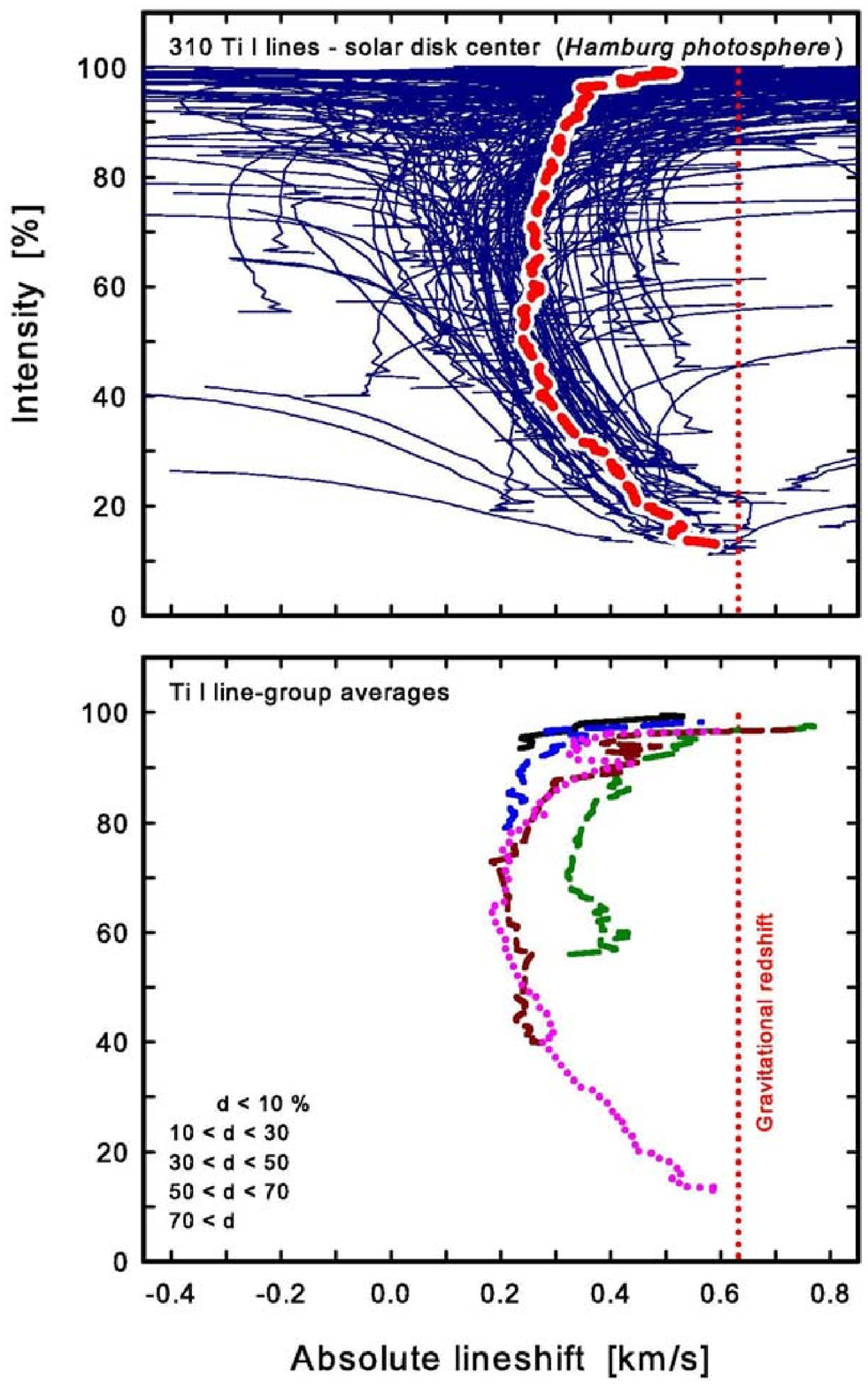}
\caption{Top: 310 \ion{Ti}{i} bisectors at solar disk center.  Solar data: Hamburg photosphere (Neckel 1999); laboratory wavelengths by Forsberg (1991). Bottom: Average bisectors for groups of lines with successively increasing line-depths $d$.}
\label{Fig4}
\end{figure}

The great number of lines, fairly well-distributed over different strengths, permits a search for line-depth dependences: Fig.~4 bottom.  The spread among individual bisectors changes with intensity level but is typically some 150~m~s$^{-1}$, requiring averaging over groups of at least some tens of lines to define a mean to significantly better than the intrinsic bisector amplitude.  This reveals a marching progression of bisector curvature with changing line-depth $d$ (becoming more horizontal for weaker lines), quite similar to what was previously known for solar \ion{Fe}{i} (Dravins et al.\ 1981; Asplund et al.\ 2000b): \ion{Ti}{i} is now the second species where this behavior is observed.  Still, for the smaller samples, effects of occasional outliers enhance the noise, and one of the curves does not follow the analogous monotonic progression in lineshift.  The number of lines in each group is $d$~$<$~10~\% -- 107 lines, 10~$\leq$~$d$~$<$~30~\% -- 79, 30~$\leq$~$d$~$<$~50~\% -- 49, 50~$\leq$~$d$~$<$~70~\% -- 29, and $d$~$\geq$~70~\% -- 46, for a total of 310.

An analogous survey was made for \ion{Ti}{ii}. Out of originally 73 candidate lines, 54 were retained for solar disk center and 46 for integrated sunlight.  Their bisector shapes in Fig.~5 resemble those of \ion{Fe}{ii} in Fig.~3, although this sample contains fewer lines (and the number is too small to permit further subdivision for different line depths).  Laboratory wavelengths originate from Zapadlik et al.\ (1995), Zapadlik (1996), and Johansson (private comm.), and are listed in Table 4 (electronic version only).

\begin{figure}
\centering
\includegraphics[width=6cm]{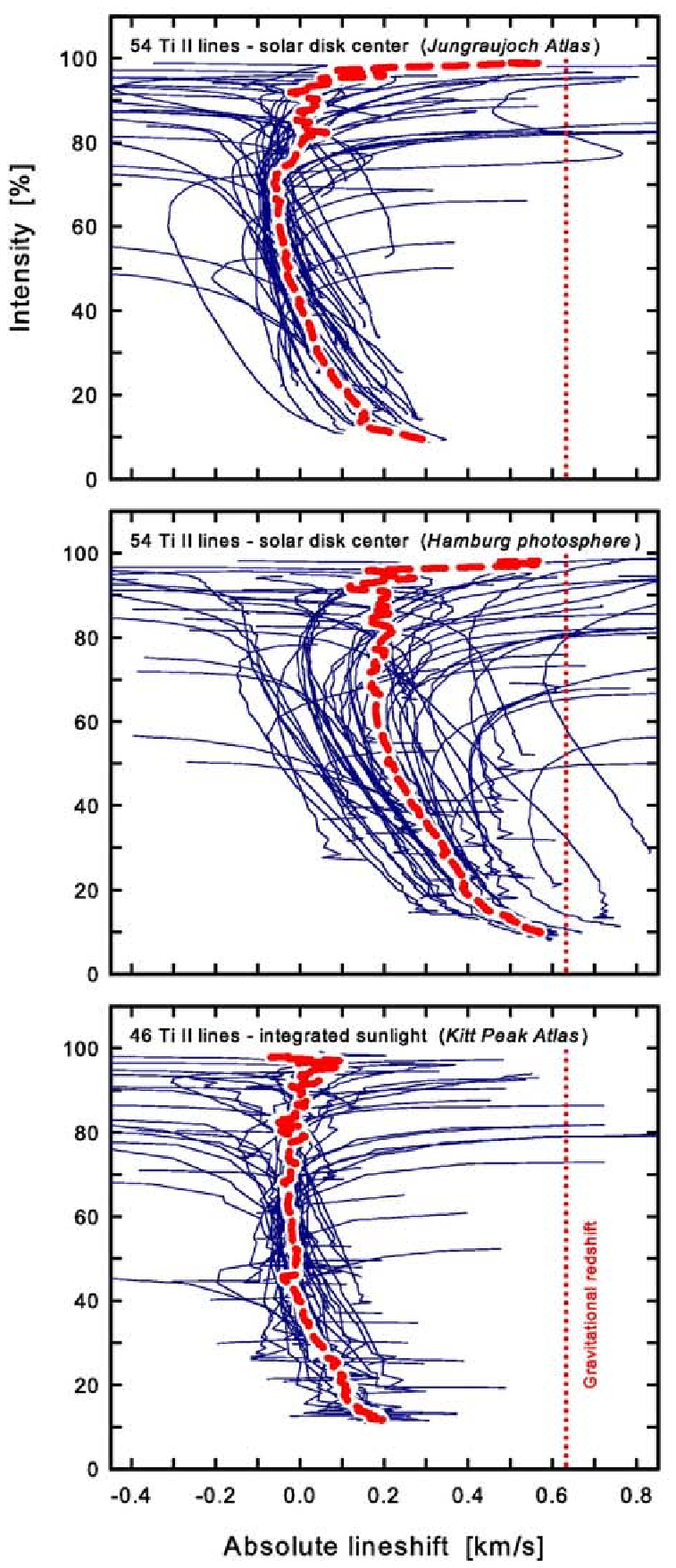}
\caption{\ion{Ti}{ii} bisectors at solar disk center from the Jungfraujoch Atlas (top) and the Hamburg photosphere (center), and in integrated sunlight from the Kitt Peak Atlas (bottom).  The data differ in average wavelength shift and in the scatter about their average.  Laboratory data: Zapadlik et al.\ (1995), Zapadlik (1996), Johansson (private comm.)}
\label{Fig5}
\end{figure}

There are somewhat bewildering differences between the two solar disk-center spectra in Fig.~5.  Their recording methods were different: a grating spectrometer for the Jungfraujoch Atlas and a Fourier Transform Spectrometer for the Hamburg data.  However, the spectral resolutions and the photometric noise are comparable and -- even if they are from different epochs and use somewhat different integrating apertures -- we would not expect significant physical differences between these recordings.  A closer examination reveals that individual bisector shapes are quite similar, although the average wavelength shifts differ, as does the scatter about the average.  Somewhat different zero-points of wavelength scales can be expected given the different calibration methods, but the origins of different spreads are less obvious.

Possibly, this could be specific to interferometric spectrometers and arise from noise in the longest sinusoidal components of the interferogram.  The spectrum is synthesized as a sum of sinusoidals where each is measured sequentially by the mechanically scanning FTS.  Phase errors for the longest components will displace whole groups of spectral lines without necessarily deforming the shapes of each of them (like a big wave moving whatever smaller-scale structures float on top of it), plausibly causing patterns such as in Fig.~5.  A possibly similar effect was noted for the Kitt Peak Atlas: the lowest frequency Fourier components are the strongest ones and are most affected by nonlinearities in the detector (Kurucz et al.\ 1984).

Another possibility is a bias from some calibration adjustment. Neckel (1999) mentions that alignment errors were corrected by comparing with the wavelength tables of Pierce and Breckinridge (1973), somehow normalizing the wavelength scale to these.  That both \ion{Ti}{i} and \ion{Ti}{ii} line-bottoms have close to zero shift may arouse suspicion that some misguided calibration may have taken place.  Even if otherwise accurate, any such tables suffer from the limitation of only one scalar number per line, not specifying to which position along each line's bisector the stated wavelength refers to.

\subsection{Solar \ion{Cr}{ii}}

Chromium is another species not studied much in this context, an iron-group element of atomic mass comparable to that of Fe or Ti, and likewise essentially free of the complicating issue of hyperfine structure.  In particular, it supplies a further sample of ionized lines, a category in which the solar spectrum is meager.

The \ion{Cr}{ii} sample is not very large.  Out of initially 57 candidates, 35 lines could be retained for disk center, and 32 for integrated sunlight (Fig.~6).  The bisector shapes are now different from previous ones, with the upper portion of the `C'-shaped bisector bending over strongly towards the red.  The lineshift differences between the atlases for disk center and integrated sunlight are analogous to those seen for other species.  Laboratory wavelengths (Johansson, private comm.) are in Table 5 (electronic version only).

Synthetic \ion{Cr}{ii} lines do not appear to have been published, and it remains to be seen if this signature of a strong redward bisector bend close to the continuum can be confirmed; however, it can be noted that rather similar lineshapes (a slight intensity depression in the redward wing) are predicted for \ion{Fe}{i} lines in hotter main-sequence A-type stars (Steffen et al.\ 2005).

\begin{figure}
\centering
\includegraphics[width=6cm]{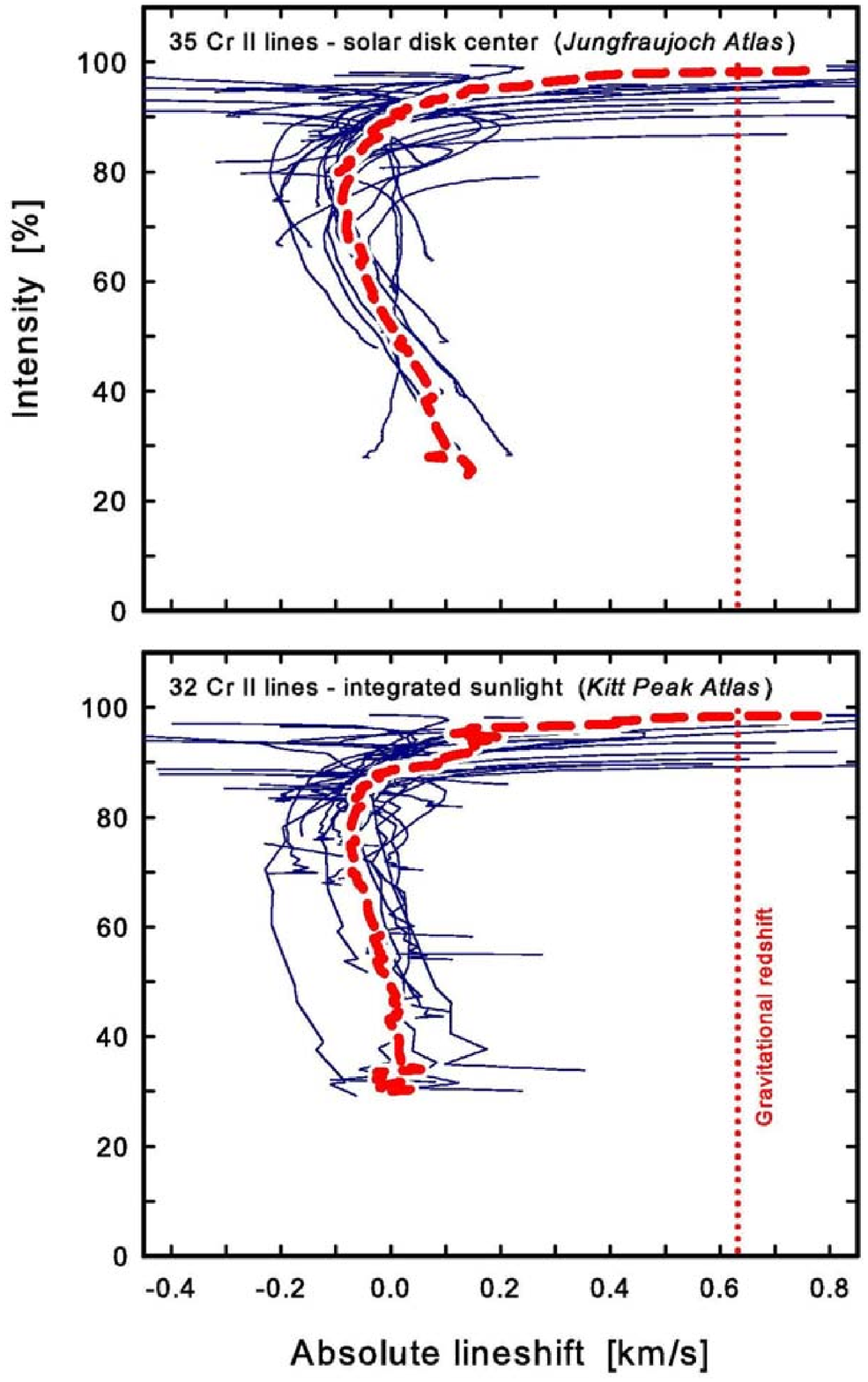}
\caption{Solar bisectors for \ion{Cr}{ii}; top -- solar disk center; bottom -- integrated sunlight.  Laboratory wavelengths are from the FERRUM Project (Johansson, private comm.)}
\label{Fig6}
\end{figure}

\subsection{Solar \ion{Ca}{i}}

Going to somewhat lighter species, we next examine \ion{Ca}{i}, of which 47 lines were identified jointly in the Sun and the laboratory.  Many of these turned out to be significantly blended, and only about half were judged clean enough to be retained: Fig.~7.  The upper parts of the bisectors are less curved than the previous ones from \ion{Cr}{ii}, but their detailed interpretation will likewise have to await detailed modeling.  Laboratory wavelengths from Litz\'{e}n (private comm.), identifying lines selected in solar and Procyon spectra, are in the electronic Table 6.

\begin{figure}
\centering
\includegraphics[width=6cm]{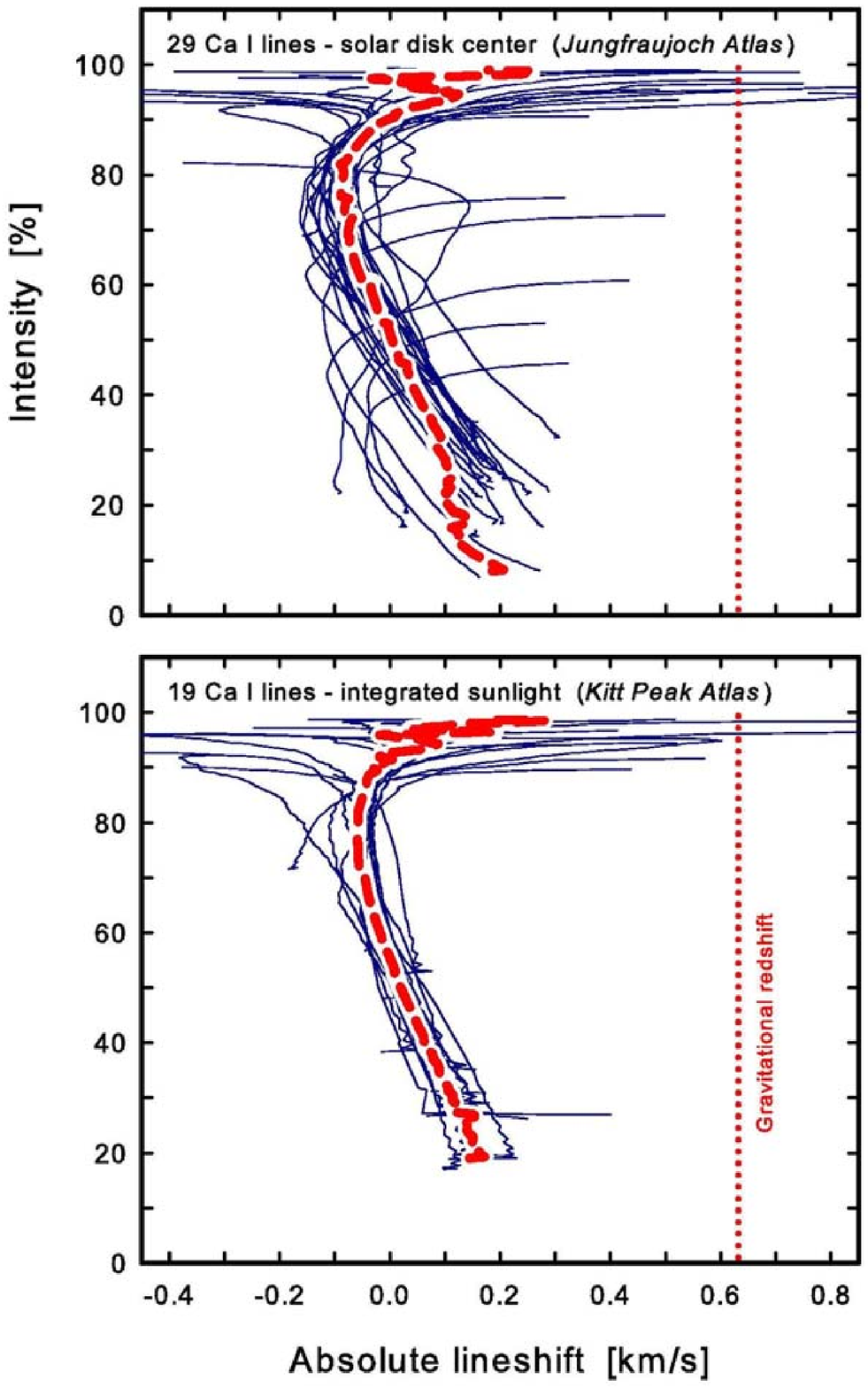}
\caption{Solar \ion{Ca}{i} line asymmetries and shifts; top -- solar disk center; bottom -- integrated sunlight.  Laboratory data by Litz\'{e}n (private comm.)}
\label{Fig7}
\end{figure}

\subsection{Weak and broad: \ion{C}{i}}

The lightest element for which photospheric line statistics can be studied is carbon.  The low atomic mass of carbon implies a significant thermal broadening but only very few \ion{C}{i} lines in the visual are strong or undisturbed enough for bisector studies (although several more may be used for abundance determinations: Lambert 1968; Asplund et al.\ 2005).  Laboratory wavelengths originate from Wiese et al.\ (1996).  Figure 8 shows \ion{C}{i} bisectors for disk center and for integrated sunlight, qualitatively similar to those of the weakest \ion{Fe}{i} or \ion{Ti}{i} lines.  The sample of lines comes from a narrow and particulary clean portion of the spectrum in the red: $\lambda\lambda$ 711.147, 711.318, 711.517/711.518, 711.699, 711.966, and 713.211~nm.

It is sobering to realize that these bisectors constitute examples of spectral features that cannot be studied yet in {\it {any}} stellar spectrum, because there the noise level is still some order of magnitude too high.  Even our solar resolutions of 500,000 and continuum S/N ratios approaching 3,000 only permit the somewhat marginal and noisy measurements in Fig.~8.

\begin{figure}
\centering
\includegraphics[width=5.85cm]{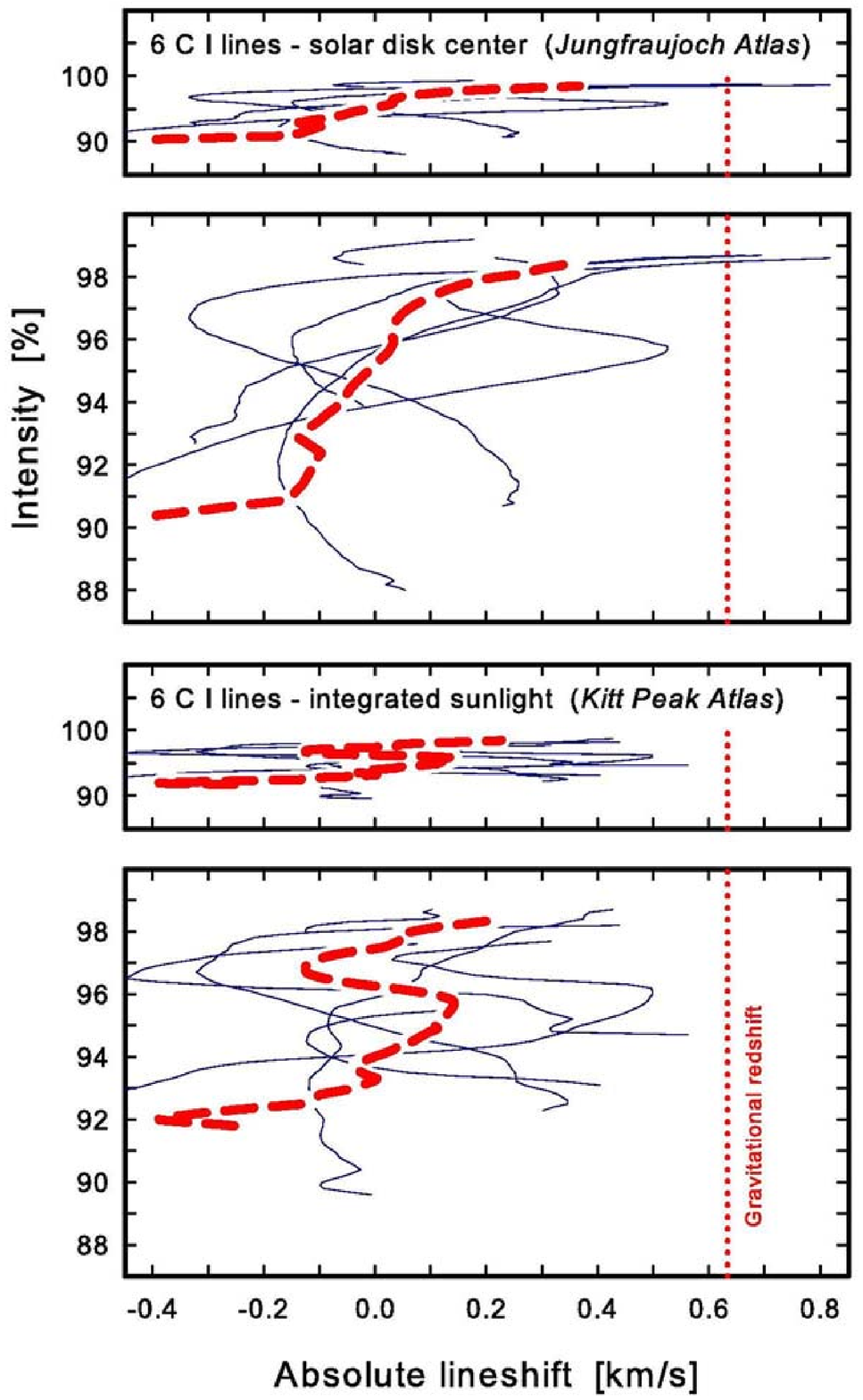}
\caption{Bisectors for the few weak `unblended' solar \ion{C}{i} lines, and their averages. The upper plots in each pair are on the same scale as other bisector figures; the bottom frames have the vertical intensity scale expanded.   Bisector measurements in broad lines and near the spectral continuum are photometrically very demanding and these types of spectral signatures are as yet not observable in any [non-solar] stellar spectrum.}
\label{Fig8}
\end{figure}

\subsection{Failures}

Given the aim of extracting spectral information to the limit, the effort must fail when some frontier is reached.  This report would therefore not be complete without mentioning at least some such `failure', i.e. a case where attempts to study lineshapes were made, but where the spread in bisectors or wavelengths made it impossible to deduce sensible values.  In Sect.~2 above, the failed attempts to extract meaningful data from the spectra of early-type, late-type, and metal-poor stars were mentioned.  However, limitations were also encountered in the analysis of highest-quality solar spectra.

One group of potentially interesting features is made up of \ion{Fe}{i} lines superposed onto the extended wings of the \ion{Ca}{ii}~H \& K absorption lines in the violet.  The atmospheric opacity at these wavelengths becomes a summation of the contributions from both \ion{Fe}{i} and \ion{Ca}{ii}, in effect lifting the formation height of the \ion{Fe}{i} lines up to much higher atmospheric levels than the weakness of the Fe lines alone would suggest.  Although synthetic line profiles do not yet appear to have been computed for such cases, their line-formation attributes must be reflected in shapes and shifts different from those in ordinary \ion{Fe}{i} lines.  On the Sun, such lines probably probe regions of `inverted' granulation where the velocity-brightness correlation is reversed relative to the deeper photosphere (Rutten et al.\ 2004; Leenaarts \& Wedemeyer-B\"{o}hm 2005; Cheung et al.\ 2007), likely leading to reversed line asymmetries (Uitenbroek 2006).  Such lines even go into emission near the solar limb (Rutten \& Stencel 1980; Watanabe \& Steenbock 1986), and in some stars, the combined opacity from such superposed lines lifts the level of formation up into the chromosphere: the hydrogen line H~$\epsilon$, superposed on the \ion{Ca}{ii}~H line wing, goes into emission in K-type stars such as Arcturus (Ayres \& Linsky 1975).

About a dozen \ion{Fe}{i} lines in the wings of each of \ion{Ca}{ii}~H and K were measured; however, the spread in lineshapes was such that no credible bisector averages could be obtained, not even for the best solar spectra.  The limitation in this case is thus `astrophysical', i.e. caused by a scarcity of suitably unblended spectral lines within the relevant wavelength interval.  For theoretical modeling, a possibility still exists (at least in principle) of computing some specific (though somewhat blended) single-line feature, although that will require detailed laboratory data not only for the target line, but also for all the blending ones.

\section{\object{Procyon} and other F-type stars}

The above discussion has demonstrated that, even for (so-called) `very high-resolution' and `very low-noise' solar spectra, bisector studies are hitting various noise limits.  In stellar cases, the spectral resolutions are much lower, yet the photometric noise is higher, often compounded by stellar rotational line broadening.  Realizing the ensuing data limits, not all atomic species were analyzed in all stars, and the often inconclusive results will not be shown here.  However, we give examples of bisector signatures in the well-observed star Procyon and in some other stars of comparable spectral types.

\subsection{\ion{Fe}{ii} in Procyon}

Ionized lines become more prominent in hotter stars, and we utilize the improved laboratory wavelengths from the FERRUM Project (Johansson 2002) to place \ion{Fe}{ii} bisectors in Procyon on a relative wavelength scale.  A limitation here, as in most stellar studies, is that these cannot be placed yet on an absolute scale since the true radial motion of the stellar center-of-mass is unknown, leaving a zero-point offset.  The exception is when stellar radial motion can be independently determined from astrometry, still only feasible for a limited number of stars (Dravins et al.\ 1999; Lindegren et al.\ 2000; Madsen et al.\ 2002).

Procyon (F5 IV-V) is a star with a particularly well-studied spectrum, available in different independent datasets, and its surface hydrodynamics and ensuing line profiles have been modeled by several groups (e.g., Atroschenko et al.\ 1989; Dravins \& Nordlund 1990b; Allende Prieto et al.\ 2002a).  However, these previous studies primarily concerned \ion{Fe}{i}, with only limited discussion of \ion{Fe}{ii}.

\begin{figure}
\centering
\includegraphics[width=6cm]{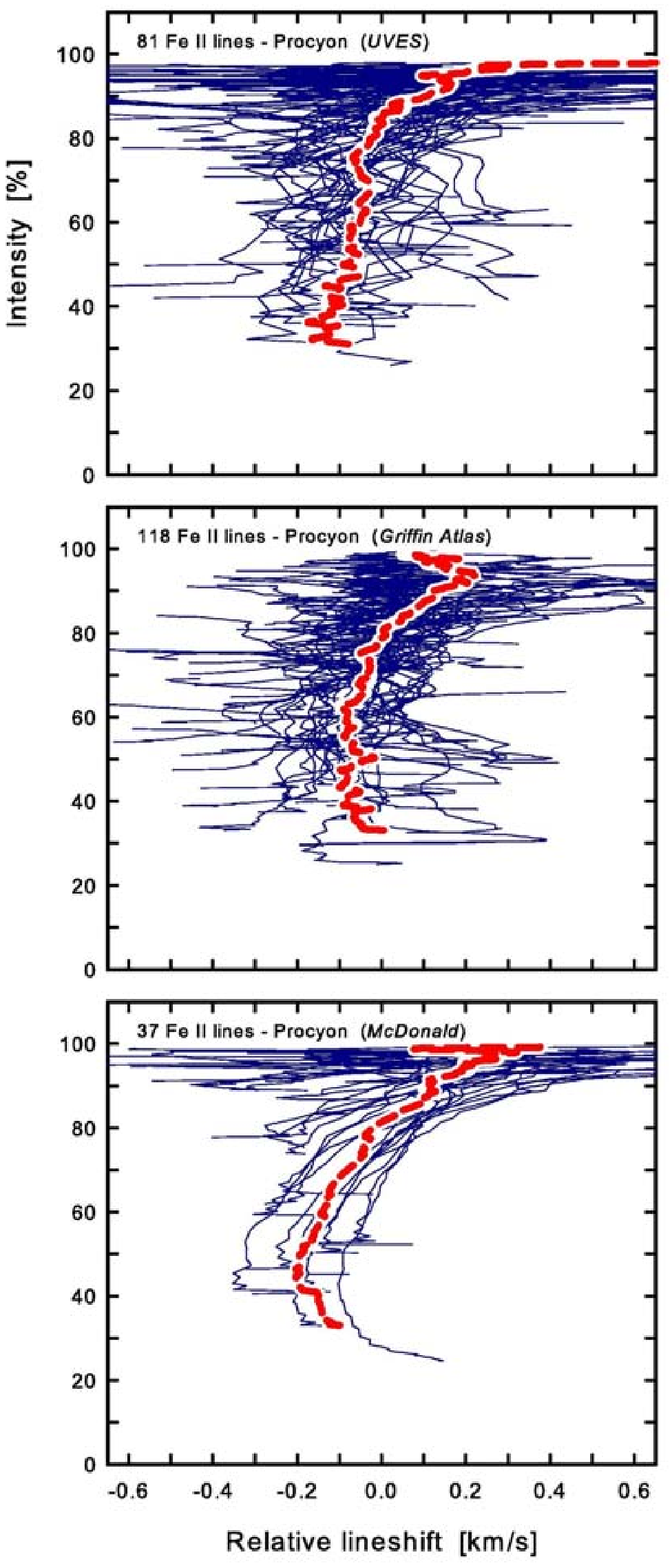}
\caption{\ion{Fe}{ii} bisectors in the F5~IV-V star Procyon, measured with successively higher spectral resolutions.  Top: 81 lines from UVES Paranal spectra (Bagnulo et al.\ 2003; R~=~80,000); Middle: 118 lines from the Procyon atlas by Griffin \& Griffin (1979; R~=~160,000); Bottom: 37 lines from McDonald, with the highest resolution (R~=~200,000) but smallest spectral coverage (Allende Prieto et al.\ 1999).  The horizontal axis only shows relative lineshifts.}
\label{Fig9}
\end{figure}

Figure 9 shows \ion{Fe}{ii} bisectors extracted from three different datasets.  The increased bisector curvature in the McDonald data certainly must be real and is achieved thanks to their higher spectral resolution.  The merit of spectral resolution can be appreciated by noting that -- given that the relevant width of a typical photospheric line may be, say, 12~km~s$^{-1}$ -- a resolution of R~=~100,000 (= 3~km~s$^{-1}$) gives 4 resolution points across that line profile, while a resolution of R~=~300,000 (= 1~km~s$^{-1}$) gives 12 points.  A bisector is obtained by averaging pairs of resolution points in the shortward and longward flanks of the line.  The number of data points on the bisector is thus half these numbers, in these examples 2 and 6, respectively.  The two points obtained in the first case can only define a straight line, indicating the bisector's wavelength position and its slope (i.e. the sense of line asymmetry).  To detect any curvature on a bisector requires at least three points on the curve, i.e. at least six photometric measurements across the relevant portions of a line.  The exact resolution required of course depends on the intrinsic width of the lines, and the requirements are lower for broader lines in F-type stars than for, e.g., solar ones.  The gradual degradation of bisector shapes caused by gradually lower resolving power (such as seen in Fig.~9) is illustrated for a Procyon model by Dravins \& Nordlund (1990b; their Fig.~10).  To nearly resolve bisector shapes in full requires resolutions comparable to those in solar atlases, by far not realized here.  The lines entering Fig.~9 are identified in the electronic Table 2. 

\subsubsection{Blueward hook of the bisectors}

Highest fidelity spectra start to reveal bisector shapes beyond their basic curvature and shift, e.g. the McDonald data in Fig.~9 suggest a sudden change in bisector behavior near continuum intensity, the curve taking a sudden turn to the blue, i.e. a `blueward hook'.  A similar signature was first seen in Procyon by Dravins (1987b), in the strongest among a small number of selected very clean lines, observed at resolutions R~$\approx$~200,000.

Such a blueward hook can already be reproduced by simple models with a multi-stream summation of wavelength-shifted line components (Dravins 1990; his Figs.~6 and 7), and be traced to the extended Lorentzian wings of the stronger, saturated, and blueshifted line components.  Their contribution in one flank of the spatially averaged line also affects the intensity in the opposite flank, in contrast to Gaussian-like components, whose absorption disappears over a short wavelength distance.  The effect is especially noticeable close to the continuum, where a small blueshifted intensity depression in the outer line flank may dominate the bisector, causing a sudden blueward hook.  Hydrodynamic modeling of Procyon spectra by Dravins \& Nordlund (1990b) did not reveal this signature because the need to extend calculations to almost the spectral continuum was not understood yet at that time.  The mechanisms were clarified in the modeling by Allende Prieto et al.\ (2002a) who found all strong Fe lines to show such a blueward hook, while not present in weaker ones.  The steeper temperature gradients in the rising and blueshifted granules produce stronger absorption lines, which therefore tend to first saturate and develop Lorentzian damping wings in those spatial locations (rather than in the redshifted and sinking intergranular lanes).

Given that this signature is visible in different datasets and is also theoretically understood, it may be accepted as real.  Can it also be seen in other F-type stars?

\subsection{Fe II in other F-type stars}

F-type stars most often have much higher rotational velocities than the solar $\approx$~2~km~s$^{-1}$ or the $\approx$~3~km~s$^{-1}$ deduced for Procyon (Allende Prieto et al.\ 2002a), generating line broadening, blending, and ensuing difficulties in bisector measurements.

Within the UVES Paranal survey, the sharpest-lined star of a spectral type close to that of Procyon (F5~IV-V) seems to be \object{${\theta}$~Scl} (HD 739, HR35) with spectral classifications in the literature going as early as F3, with the most recent F5~V.  Its rotational velocity does not appear to have been measured, but a visual comparison of its line profiles to those of Procyon suggests a very comparable value of $Vsini$.  Edvardsson et al.\ (1993) deduced a metallicity [Fe/H] = --0.10 (comparable to the Procyon value of --0.05; Allende Prieto et al.\ 2002a), and the surface gravity log g [cgs]= 4.26, higher by $\approx$~0.3 dex than for Procyon's luminosity class IV-V, placing this star on the main sequence.

The \ion{Fe}{ii} lines were examined in ${\theta}$~Scl, analogous to the case for Procyon, but here more lines could be identified, retained, and measured close to the continuum.  The individual bisectors in Fig.~10 have a considerable spread, but their average does display a sudden `blueward hook' close to the continuum, not unlike Procyon.  The limited spectral resolution hides signatures of bisector curvature for intensity levels below some 80\% (where the number of spectral resolution elements across the relevant line-width is too low), but does not constrain the wavelength position of the bisector close to the continuum, where the line wings are broader and the number of resolution elements thus is correspondingly greater.  The lines selected for Fig.~10 were in the electronic Table 2.

\begin{figure}
\centering
\includegraphics[width=6cm]{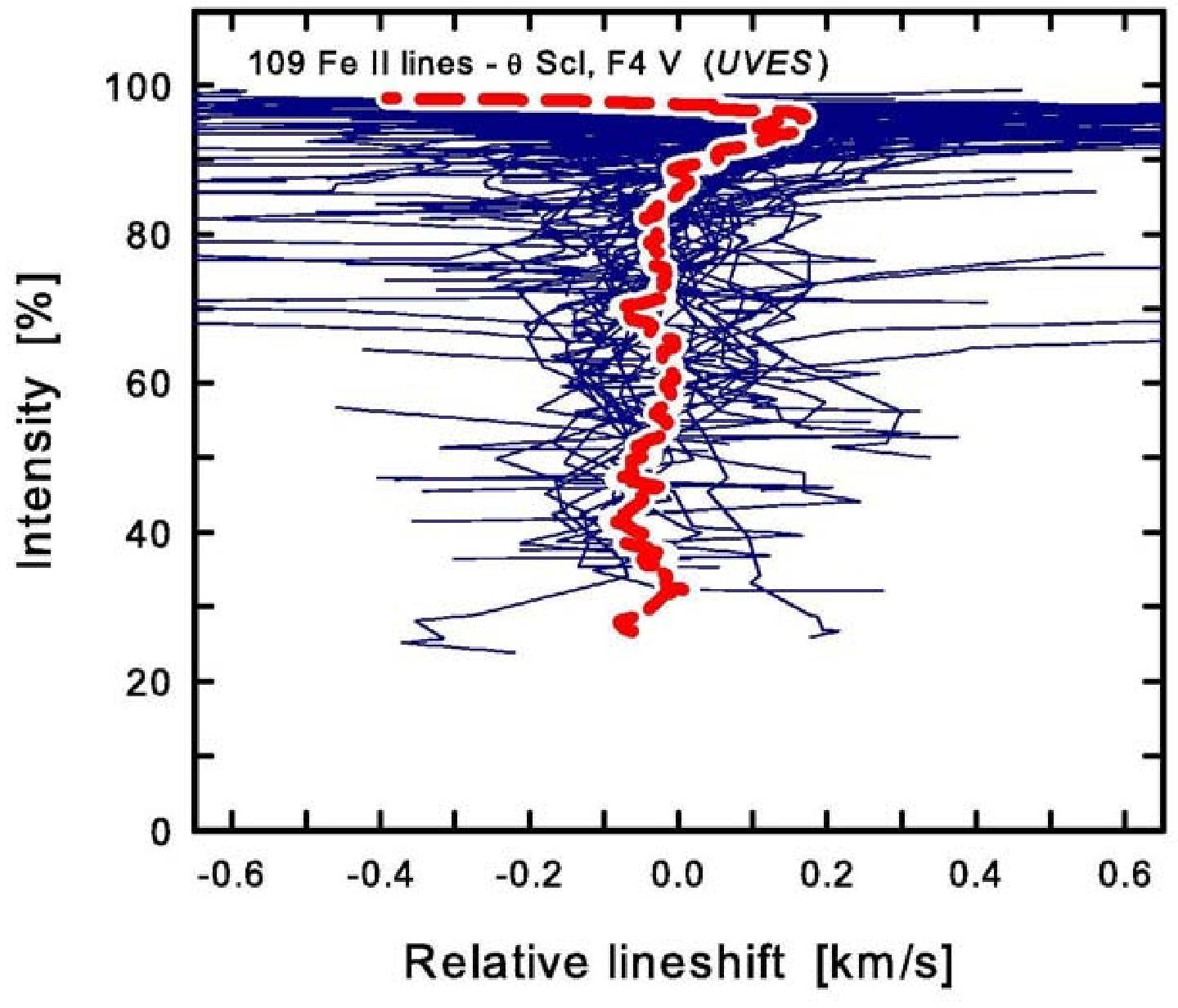}
\caption{\ion{Fe}{ii} line bisectors in the F5~V star ${\theta}$~Scl (from UVES Paranal data), showing a sudden `blueward hook' of its average bisector very close to the continuum.}
\label{Fig10}
\end{figure}
  
Accepting these F-star signatures as real demonstrates how different degrees of line saturation across stellar surface inhomogeneities can be detected in integrated starlight, also permitting tests of rather detailed properties of hydrodynamic models.  However, we are not far from the limits in extracting spectral-line signatures from spectroscopy of integrated starlight.  To detect these signatures requires stars with both astrophysically benign spectra (very little rotational broadening), high-resolution instrumentation, very low-noise recordings, followed by the averaging over some hundred spectral lines.

\subsection{Other species: \ion{Ti}{i} and \ion{Ca}{i}}

Next to iron, the species represented with the greatest number of lines is titanium, where 230 \ion{Ti}{i} lines could be retained in Procyon UVES spectra.  This significant number permits a meaningful bisector average to be computed: Fig.~11.  However, note that (as opposed to solar data and the analogous Fig.~4) we no longer have any absolute lineshifts, only relative ones.

If the spectrometer wavelength scale is precise enough, one can use its wavelengths relative to the laboratory ones to get relative lineshifts with a similar (although unknown) offset.  However, such an attempt resulted in a clearly unphysical bisector spread not only among individual bisectors but also between line-depth averages, precluding analyses of such a line-depth dependence.  The reasons are unclear but are unlikely to be due to laboratory wavelength issues, given the smoother lineshift patterns of largely the same lines in solar spectra (Fig.~4).  It must also be noted that, although the total line sample is substantial, its distribution of line-strengths is fairly uneven (most lines are weak), much worse than for the \ion{Fe}{ii} sample in Fig.~9.  Instead, the wavelength of each bisector was first individually averaged to zero, and the average bisector shape then calculated (Fig.~11 top).  However, this procedure must decrease the bisector curvature somewhat since only the shapes are averaged, not accounting for different shifts between differently strong lines.

Still, there is again (at least slight) a suggestion of the blueward bend of the bisector close to the continuum.  However, the measurement is more marginal than for \ion{Fe}{ii} since the number of strong lines contributing to this signature is fewer.
 
A somewhat lighter atomic species is calcium.  Although \ion{Ca}{i} is represented by much fewer lines, many of these are relatively strong and contribute to a better-defined bisector.  The much smaller spread in line-depth produces a rather well-defined average (Fig.~11 bottom).  The lines selected for Fig.~11 are listed in the electronic Tables 3 and 6 for \ion{Ti}{i} and \ion{Ca}{i}, respectively.

\begin{figure}
\centering
\includegraphics[width=6cm]{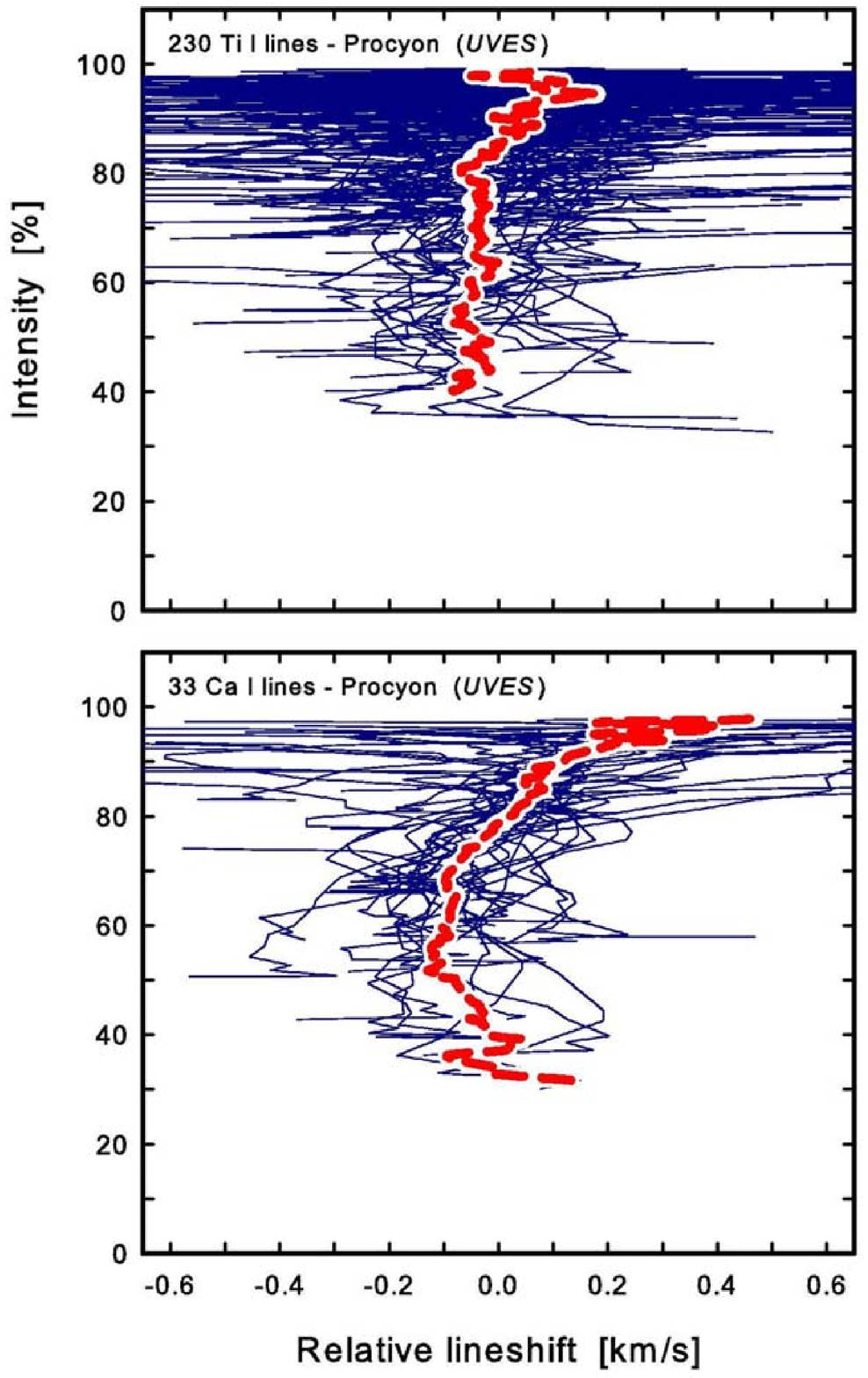}
\caption{\ion{Ti}{i} and \ion{Ca}{i} line bisectors in Procyon (UVES Paranal spectra).  Among the non-iron species, titanium is the one with the richest number of lines.  Top: Individual bisectors and their average for 230 \ion{Ti}{i} lines; Bottom: 33 \ion{Ca}{i} bisectors and average.}
\label{Fig11}
\end{figure}

\section{Frontiers in high-fidelity spectroscopy}

Having now prodded the limits of current spectroscopy, we now consider to what extent the limiting factors are understood and whether they can be circumvented or remedied, especially in view of planning future facilities.

\subsection{Adequate spectral resolution?}

Some aspects are `straightforward' (at least in principle), such as the need for improved spectral resolution.  The damaging effects on bisector shapes and shifts caused by lesser resolutions are obvious in both observations (Fig.~9) and in simulations (Dravins 1987a; Dravins \& Nordlund 1990b).  If the instrumental profile of the spectrometer is well known, that of course can be convolved with the output from theoretical line-profile modeling to enable a comparison with observations.  However, while [de]convolution techniques may permit  analyses of smeared-out line profiles, they cannot increase the spectral resolution beyond the level at which the spectrum originally was sampled.  As already pointed out, any bisector point follows from measurements in opposite flanks of the line; a bisector shape to be defined by five points, say, requires ten positions, which for photospheric lines in solar-type stars implies resolutions no lower than R~$\approx$~200,000.

However, grating spectrometers of very high resolution are non-trivial for matching to large telescopes, and a desired high light efficiency could demand adaptive optics (Ge et al.\ 2002; Sacco et al.\ 2004), not simple at shorter wavelengths.  Fourier transform spectrometers give adequate resolution, but in broad-band stellar observations they are seriously limited by photon noise.

While some specialized instruments for really high-resolution spectroscopy (R~$\approx$~10$^{6}$) have been used to measure interstellar lines (e.g., the Ultra-High-Resolution Facility at the Anglo-Australian Telescope; Diego et al.\ 1995), the only seriously high-resolution night-time instrument with extended spectral coverage currently foreseen appears to be PEPSI for the Large Binocular Telescope, designed for resolutions up to R~=~300,000 (Strassmeier et al.\ 2003).  Among the PEPSI science cases (Strassmeier et al.\ 2004), diagnosing stellar hydrodynamics is examined, concluding -- similar to the present discussion -- that studies of bisector curvature require resolutions in excess of $\approx$~300,000 and cross-dispersed echelle spectra, so that the noise inherent in any one spectral line can be overcome through the statistics of many. 
Steffen \& Strassmeier (2007) describe the PEPSI `deep spectrum' project for providing the highest quality optical spectra for any star other than the Sun.  The goal is a signal-to-noise ratio of 5,000:1 at a spectral resolution of 1 km~s$^{-1}$ covering the entire optical spectrum.  Certainly this is a worthy goal, although, at such levels, the accuracy may well be limited by systematic errors, not statistical noise.  Even the highest-resolution solar atlases have been contaminated by various subtle effects such as light leaks in order-separating color filters (e.g., Dravins 1994).

\subsection{Avoiding telluric absorption?}

Even with superlative spectral resolution and photometric precision, spectra are distorted by telluric absorption and emission lines and bands, compromising the spectral fidelity.  Actually, in some wavelength regions, the level of contamination makes it doubtful whether it will ever be possible to record truly high-fidelity solar or stellar spectra from the ground.

In some regions, telluric lines from, say, H$_{2}$O, are sharp and well confined in wavelength, and then can be identified and removed in the analysis (e.g., Hadrava 2006).  However, an accurate removal requires the terrestrial lines to be spectrally resolved, and not even the `high' resolution in solar atlases is fully adequate here.  Inspection of telluric lines in clean spectral regions of the Kitt Peak Atlas (for an example, see $\lambda$~1250.7 nm) shows that the lines are still unresolved, instead displaying the characteristic `ringing' instrumental profile of an FTS, even at the resolution R~=~500,000.

There are other and more treacherous telluric features, extending over wider wavelength regions.  These include diffuse absorptions due to ozone O$_{3}$ and the oxygen dimer [O$_{2}$]$_{2}$ that appear throughout the visual. Ozone produces the atmospheric transmission cutoff in the near ultraviolet, where old stellar spectrograms can actually be used to infer past amounts of terrestrial ozone (Griffin 2005).

Only in a few cases has it been possible to compare high-resolution spectra recorded from above and beneath the atmosphere.  Sirius was observed with the Goddard High-Resolution Spectrograph on the Hubble Space Telescope in the ultraviolet up to 320 nm, while ground-based observations reach down to 305 nm.  The overlapping interval is very instructive in demonstrating how the apparent continuum in ground-based recordings can be systematically wrong by some 10 \% due to diffuse telluric absorptions (Griffin 2005, and private comm.).

Kurucz (2005) computed synthetic telluric spectra, including O$_{3}$ and [O$_{2}$]$_{2}$, showing how they affect the highest-resolution solar spectra.  The richness of the telluric spectra means that weak lines with depths of perhaps only some percent or less are often superposed onto the flanks or cores of solar ones, making the reduction for their effects awkward or practically impossible (especially since atmospheric extinction is variable in time; Stubbs et al.\ 2007).  In fact, for many spectral regions, this sets a signal-to-noise limit on the order of 100 or worse, irrespective of the photometric precision reached.  The affected regions can be inspected in a spectral atlas of telluric absorption (R~$\approx$~300,000), derived from solar data by Hinkle et al.\ (2003).

Observations from orbit would eliminate telluric absorptions (although geocoronal emission would still be present), but any space mission for high-resolution stellar spectroscopy would probably be complex, considering not only that a large telescope is needed but also that the Doppler shift induced by the spacecraft motion implies continuous wavelength changes.

\subsection{Accurate wavelength calibration?}

Grating and Fourier transform instruments establish their wavelength scales through different schemes, and each is likely to have its characteristic signatures in its wavelength noise.

Grating spectrometers commonly use an emission-line lamp (thorium, etc.), and thus their wavelength scale depends upon the accuracy of the corresponding laboratory wavelengths and on how well the lamp at the telescope reproduces the laboratory sources.  Traditionally, the wavelength noise among individual lines may have been around 100~$\mathrm{m\, s^{-1}}$ but recent calibrations of thorium-argon hollow-cathode lamps have now decreased the internal scatter to some 10 $\mathrm{m\, s^{-1}}$ (Lovis \& Pepe 2007), further permitting a selection of subsets of lines for best calibration (Murphy et al.\ 2007a).  

Despite good internal precision, the external link to absolute wavelengths may still be considerably worse. Some estimates of external noise are possible by comparing with the solar lineshifts themselves, assuming they are adequately reproduced by existing models.  While this might appear to contradict the primary purpose of using wavelength shifts to calibrate models (rather than the opposite), one may use the fact that different classes of lines (with different amount of predicted shift) are intermingled in wavelength and present a quite complex pattern of differential wavelength displacements.  If a similar specific pattern is visible in the observed spectra, that must give some credibility to the spectrometer accuracy.

Fourier transform spectrometers have their specific issues.  The interferogram is recorded sequentially for its sinusoidal components, between low Fourier frequencies (sinusoidal components of the spectrum) and high frequencies (how high determines the spectral resolution).  However, the realities of detectors and photon noise normally preclude the entire spectrum being measured at once; instead only one piece at a time must be selected by some pre-filter or similar device.  The transmission functions of these pre-filters have to be precisely calibrated to enable a correct continuum level to be set.  Its placement to better than 1:100 is no trivial task, as illustrated by the efforts to improve the continuum in the Kitt Peak atlases by Neckel (1999) and Kurucz (2005). 

Other subtle issues exist for solar spectra: `solar-disk center' is not a uniquely defined location relative to the solar rotational axis and its activity belts (given the 7-degree inclination of the solar axis from the perpendicular relative to the plane of the ecliptic confining an Earth-bound observer); the size and shape of the sampling area at the center of the Sun measurably affect the line shapes due to different ways of averaging; different integration times yield different amounts of line-smearing due to p-mode oscillations; and even for integrated sunlight (which is also not simple to obtain) there are subtle variations during the solar 11-year activity cycle.

The Kitt Peak Atlas of integrated sunlight was pieced together from eight FTS scans, somewhat overlapping in wavelength, each of which was integrated for two hours, so the time resolution is two hours for a given scan (different sections were recorded on different days).  A side effect is a spectral broadening of some 200 m~s$^{-1}$, caused by the changing Sun-Earth radial velocity during each integration.

For the wavelength calibration of interferometric spectrometers, laboratory laser lines are often used (most often He-Ne $\lambda$~628 nm), albeit with the problem that it is next to impossible to ensure identical light-paths inside the instrument for both the astronomical and the calibration spectra (and thus the wavelength scales become displaced).  To circumvent this issue, the Kitt Peak Atlas of integrated sunlight instead used one very sharp telluric O$_{2}$ line as reference.  Since this is superposed onto the solar spectrum, light-path differences are avoided, and the wavelength scale of the entire atlas was set by applying its laboratory wavelength (Kurucz et al.\ 1984).  Still, since that particular line only appears in two of the eight FTS scans, the other spectral segments had to be fitted by the overlapping of adjacent ones.

However, the O$_{2}$ wavelength used to set the wavelength scale -- $\lambda$~688.38335 nm in standard air -- was the value accepted as the most accurate at the time the atlas was produced (stemming from measurements by Pierce \& Breckinridge 1973), but its last decimal has since been subject to readjustment.  On our desired precision levels, even the sharpest telluric lines are shifted, are asymmetric, do depend on airmass, and change over time.  Pressure shifts cause the absorption coefficient to depend on atmospheric height while the whole telluric-line bisector is displaced by wind-induced Doppler shifts (Caccin et al.\ 1985).  

Another circumstance arose for the Kitt Peak Atlas because of limitations in data processing at the time of the observations.  During the hours of integration, numerous scans across successive Fourier frequencies were co-added into a single interferogram file, from which the spectrum was later calculated.  During this integration, the airmass of observation gradually changed, as did the strength of the atmospheric O$_{2}$ line, which ended up slightly asymmetric, thus adding to the wavelength uncertainty (Kurucz, private comm.).  In the atlas text (Kurucz et al.\ 1984), it is estimated that the final wavelengths could have errors as large as 0.1 km~s$^{-1}$, even if the internal precision of each FTS scan is better.  

\section{Conclusions and solutions}

Following the above discussions, we may better appreciate the information content in spectroscopic data close to current limits in photometric and wavelength accuracy.  After surveying many solar lines, it has been possible to extract data on wavelength-shifted line bisectors from several atomic species that previously have not been modeled, and subtle bisector signatures seen in F-type spectra have been (at least marginally) confirmed.

On the other hand, systematic differences can be seen in apparent wavelength shifts between groups of lines from different species, and between samples of the same lines extracted from different spectral atlases.  With the caveat that these are still awaiting hydrodynamic modeling (and there could perhaps be some nontrivial 3-D or non-LTE effects?), such differences seem more likely due to zero-point errors and wavelength noise.  The current boundary for absolute lineshift-studies seems to be not much better than 100~m~s$^{-1}$, limited both by the quality of laboratory data and by instrumental calibrations.  Relative shifts between different line-groups can be measured somewhat better, perhaps to 30~m~s$^{-1}$.  However, credible lineshifts on these levels cannot be deduced from individual lines, but require an averaging over many, in order to reduce the astrophysical noise due to weak blends.  Some single-line profiles and shifts that have been published in the past, claiming a very close agreement with theoretical prediction, may well have been fortuitous.

Similar limitations must apply to other studies of tiny wavelength shifts, e.g., those that are being sought in quasar spectra in searches for possible changes in fundamental physical constants over cosmological time and space.  The accuracies aimed at are comparable to those we are discussing, and the error sources must be analogous.

\subsection{Future high-fidelity spectroscopy?}

Although several parameters limiting highest-fidelity spectroscopy can be identified, their immediate remedies cannot.  Avoiding the terrestrial atmosphere requires ambitious space missions, and improving laboratory wavelengths requires dedicated long-term efforts.  Possibly, the least difficult part is to improve instrumental calibrations.

Laboratory devices, in particular the laser frequency comb, permit remarkably precise wavelength determinations, and are also envisioned for future astronomical spectrometers (Araujo-Hauck et al.\ 2007; Murphy et al.\ 2007b; Steinmetz et al.\ 2008).  However, it is not necessarily the stability of the calibration device that limits the accuracy in astrophysical spectra because it may in addition require identical light paths for the source and its calibration, and without the calibration light contaminating the source spectrum.

The telescope-spectrometer interface is another issue.  Analyses of the UVES spectrometer during asteroid observations reveal a noise of typically 10--50~m~s$^{-1}$, apparently caused by a non-uniform and variable illumination in the image projected by the telescope onto the spectrometer entrance slit (Molaro et al.\ 2008).  Given the chromatic nature of atmospheric dispersion, the wavelength dependence of such shifts could also mimic line-depth dependences since stronger lines often occur at shorter wavelengths.  Although a more uniform illumination is provided by image slicers or fiber-optics feeds, their use becomes awkward if -- as in the present work -- we require both extended spectral coverage and the use of (very) large telecopes.  For non-solar work, the former requires multi-order echelle spectrometers (such as UVES), and the latter implies (very) large image scales.  Light from all image slices, projected onto the focal plane, would overlap adjacent echelle orders (precluding the recording of extended spectral regions), and an optical fiber would need to have a large diameter to embrace most starlight, requiring entrance apertures that are too wide for adequate spectral resolution (or else cause severe light losses).  Possible solutions that avoid the construction of huge instruments could include spectrometers with adaptive optics, as mentioned in Sect.~6.1.

A stable wavelength scale enables the physical flicker of stellar wavelengths to be monitored, a signal of stellar surface variability that is not only interesting in itself, but that must be understood for identifying small exoplanets from stellar barycentric radial-velocity variations.  Also the astrophysical noise of line-blends and rotationally smeared profiles could ultimately be circumvented.  This will require what could become a next major step in stellar studies, namely high-resolution spectroscopy across spatially resolved stellar disks, utilizing optical interferometers and the extremely large telescopes of the future.

Hydrodynamic models already predict the revealing behavior of absorption-line wavelengths across stellar disks, differing significantly among various stars and indicating the level of `corrugation' on their respective surfaces (Dravins \& Nordlund 1990a; Dravins et al.\ 2005).  In stars with `smooth' surfaces (in the optical-depth sense), one expects convective blueshifts to decrease from stellar disk center towards the limb, since vertical convective velocities then become perpendicular to the line of sight, and the horizontal velocities that contribute Doppler shifts appear symmetric.  However, stars with `corrugated' surfaces with `hills' and `valleys' should show the opposite, i.e. an increasing blueshift towards the limb.  There, one will predominantly see the approaching (thus blueshifted) velocities on the slopes of those `hills' that are facing the observer.   The receding and redshifted components remain hidden behind the `hills', and an enhanced blueshift results.

More effects will appear in time variability across stellar disks: On a `smooth' star, temporal fluctuations are caused by the random evolution of granules, all of which are reached by the observer's line of sight.  On a `corrugated' star, observed near its limb, another element of variability is added because the changing `corrugation' of the swaying stellar surface sometimes hides some granules from direct view.  The result is an enhanced temporal variability of line wavelengths and bisectors, constituting another observational parameter for stellar hydrodynamics.  The study of such phenomena may require spectrometers with integral-field units and adaptive optics on extremely large telescopes but promise to keep studies of precise wavelength shifts as a rewarding topic in the future.

\begin{acknowledgements}
      Many thanks are due to groups in laboratory spectroscopy who provided various line wavelengths in advance of publication, in particular the Atomic Astrophysics group at Lund Observatory where data on \ion{Ca}{i} were provided by Ulf Litz\'{e}n, and on Cr I, Cr II, Fe II, and \ion{Ti}{ii} by Sveneric Johansson.  The \ion{Co}{i} wavelengths were kindly provided by Juliet Pickering of Imperial College, London, and laboratory data for \ion{Si}{i} by Rolf Engleman of University of New Mexico, Albuquerque.   

The study used data from the UVES Paranal Observatory Project of the European Southern Observatory (ESO DDT Program ID 266.D-5655).

It also used solar spectral atlases obtained with the Fourier Transform Spectrometer at the McMath/Pierce Solar Telescope situated on Kitt Peak, Arizona, operated by the National Solar Observatory, a Division of the National Optical Astronomy Observatories.  NOAO is administered by the Association of Universities for Research in Astronomy, Inc., under cooperative agreement with the National Science Foundation.

This research has made use of the SIMBAD database, operated at the CDS, Strasbourg, France, and of NASA's Astrophysics Data System.

An early exploration of bisectors in the spectrum of Procyon was made in a Master's thesis by Thomas Hakewill (2003), carried out jointly at Lund Observatory and the University of Cardiff, Wales.

Parts of this paper were written during a stay as Scientific Visitor at the European Southern Observatory in Santiago de Chile.  Ulf Litz\'{e}n and Sveneric Johansson of Lund Observatory provided valuable comments on the manuscript.

Finally, thanks are due to an anonymous referee for very detailed comments that helped clarify a number of points.

\end{acknowledgements}

%-------------------------------------------------------------
%     For the online material, table longer than a single page
%                 In the preamble, use: \usepackage{longtable}
%       or for landscape option: \usepackage{longtable,lscape}
%-------------------------------------------------------------

% Table will be print automatically at the end, in the section Online material.

\onllongtab{2}{
\begin{longtable}{ccccccc}
\caption{\ion{Fe}{ii} spectral lines examined in various spectra and selected for the plots. X denotes accepted lines; O rejected ones, and N denotes no data.  Two lines (X) form a close doublet.  Laboratory wavelengths in air are from the ongoing FERRUM Project (Johansson 2002, \& private comm.)}\\
\hline
\hline
$\lambda$ [nm] & Solar disk center & Integrated sunlight & \object{Procyon} UVES & Procyon Griffin & Procyon McDonald & \object{$\theta$~Scl} UVES
\\
\hline
\endfirsthead
\caption{Continued.} \\
\hline
$\lambda$ [nm] & Solar disk center & Integrated sunlight & Procyon UVES & Procyon Griffin & Procyon McDonald & $\theta$~Scl UVES
\\
\hline
\endhead
\hline
\endfoot
\hline
\endlastfoot

401.84866	&	X	&	X	&	O	&	X	&	N	&	X	\\
402.45502	&	X	&	X	&	X	&	X	&	N	&	X	\\
408.72770	&	X	&	X	&	O	&	X	&	N	&	X	\\
408.87493	&	X	&	X	&	O	&	X	&	N	&	X	\\
411.95199	&	X	&	X	&	O	&	X	&	N	&	X	\\
412.26595	&	X	&	X	&	O	&	X	&	N	&	X	\\
412.47839	&	X	&	X	&	O	&	X	&	N	&	X	\\
412.87398	&	X	&	O	&	O	&	X	&	N	&	O	\\
415.17905	&	O	&	O	&	O	&	X	&	N	&	X	\\
417.34518	&	O	&	O	&	O	&	O	&	N	&	O	\\
417.88547	&	O	&	O	&	X	&	X	&	N	&	O	\\
418.31895	&	X	&	X	&	O	&	X	&	N	&	X	\\
423.31625	&	O	&	O	&	O	&	X	&	N	&	X	\\
425.81497	&	O	&	O	&	O	&	X	&	N	&	X	\\
427.33208	&	O	&	O	&	O	&	O	&	N	&	O	\\
429.65662	&	O	&	O	&	O	&	O	&	N	&	O	\\
430.31708	&	O	&	O	&	O	&	X	&	N	&	O	\\
431.43030	&	O	&	O	&	O	&	X	&	N	&	X	\\
433.86973	&	O	&	O	&	O	&	X	&	N	&	O	\\
435.17627	&	O	&	O	&	O	&	O	&	N	&	O	\\
436.12516	&	O	&	O	&	O	&	X	&	N	&	X	\\
436.94023	&	X	&	O	&	O	&	X	&	N	&	O	\\
438.43138	&	O	&	O	&	X	&	X	&	N	&	X	\\
438.53779	&	X	&	O	&	O	&	O	&	N	&	O	\\
441.35920	&	O	&	O	&	X	&	X	&	N	&	X	\\
441.68196	&	X	&	X	&	X	&	X	&	N	&	X	\\
443.91253	&	O	&	O	&	X	&	X	&	N	&	X	\\
444.62429	&	X	&	X	&	O	&	X	&	N	&	X	\\
447.29245	&	X	&	X	&	O	&	X	&	N	&	X	\\
448.91760	&	O	&	O	&	X	&	O	&	N	&	O	\\
449.13988	&	O	&	O	&	X	&	X	&	N	&	X	\\
450.82807	&	X	&	X	&	O	&	X	&	N	&	O	\\
451.53333	&	O	&	O	&	X	&	X	&	N	&	O	\\
452.02199	&	X	&	X	&	X	&	X	&	N	&	X	\\
452.26280	&	O	&	O	&	O	&	O	&	N	&	X	\\
453.41603	&	X	&	O	&	O	&	X	&	N	&	X	\\
454.15161	&	X	&	O	&	O	&	X	&	N	&	X	\\
454.94667	&	O	&	O	&	O	&	X	&	N	&	O	\\
455.58877	&	X	&	O	&	O	&	O	&	N	&	X	\\
457.63334	&	X	&	X	&	X	&	X	&	X	&	X	\\
458.28305	&	X	&	O	&	X	&	X	&	X	&	X	\\
458.38301	&	O	&	O	&	O	&	O	&	O	&	O	\\
460.13730	&	X	&	X	&	X	&	X	&	O	&	O	\\
462.05129	&	O	&	O	&	X	&	X	&	X	&	X	\\
462.93326	&	O	&	O	&	X	&	X	&	X	&	X	\\
463.53180	&	X	&	X	&	X	&	X	&	X	&	X	\\
465.69762	&	X	&	X	&	O	&	X	&	O	&	X	\\
466.11821	&	X	&	X	&	O	&	X	&	O	&	X	\\
466.37045	&	X	&	O	&	O	&	X	&	O	&	O	\\
466.67502	&	X	&	X	&	O	&	X	&	O	&	X	\\
467.01712	&	X	&	O	&	O	&	X	&	O	&	X	\\
472.01409	&	X	&	X	&	X	&	X	&	X	&	X	\\
473.14472	&	X	&	X	&	X	&	X	&	X	&	X	\\
481.82388	&	X	&	X	&	O	&	X	&	O	&	X	\\
482.57266	&	X	&	X	&	X	&	X	&	X	&	X	\\
483.31928	&	X	&	X	&	O	&	X	&	O	&	X	\\
483.99878	&	X	&	X	&	X	&	X	&	X	&	O	\\
486.87973	&	X	&	X	&	X	&	X	&	X	&	X	\\
489.38144	&	X	&	X	&	O	&	X	&	O	&	X	\\
492.39217	&	X	&	X	&	X	&	X	&	X	&	X	\\
499.33515	&	X	&	X	&	X	&	X	&	X	&	X	\\
500.07314	&	X	&	X	&	X	&	X	&	X	&	X	\\
501.84370	&	O	&	O	&	O	&	O	&	O	&	O	\\
501.94648	&	X	&	X	&	X	&	X	&	X	&	X	\\
503.69126	&	X	&	X	&	X	&	X	&	X	&	X	\\
510.06570	&	X	&	X	&	X	&	X	&	N	&	X	\\
512.78614	&	X	&	X	&	X	&	X	&	X	&	X	\\
513.26635	&	X	&	X	&	X	&	X	&	X	&	X	\\
513.67953	&	X	&	X	&	X	&	X	&	X	&	X	\\
514.61189	&	O	&	O	&	X	&	X	&	X	&	X	\\
515.44025	&	O	&	O	&	X	&	X	&	X	&	X	\\
516.08410	&	X	&	X	&	X	&	X	&	X	&	X	\\
516.11754	&	X	&	X	&	X	&	X	&	X	&	X	\\
516.90280	&	O	&	O	&	O	&	O	&	O	&	O	\\
519.75705	&	X	&	X	&	X	&	X	&	X	&	X	\\
523.46246	&	X	&	X	&	X	&	X	&	X	&	X	\\
525.69328	&	X	&	O	&	X	&	X	&	N	&	X	\\
526.48039	&	X	&	X	&	X	&	X	&	X	&	X	\\
527.59982	&	O	&	O	&	X	&	X	&	O	&	X	\\
528.41036	&	X	&	X	&	X	&	X	&	X	&	X	\\
531.66096	&	O	&	O	&	O	&	O	&	O	&	O	\\
531.67835	&	O	&	O	&	O	&	X	&	O	&	X	\\
532.55552	&	X	&	X	&	X	&	X	&	X	&	X	\\
533.77235	&	O	&	O	&	X	&	X	&	N	&	X	\\
534.65644	&	O	&	O	&	X	&	X	&	X	&	X	\\
536.28616	&	O	&	O	&	X	&	X	&	O	&	O	\\
540.88149	&	X	&	X	&	X	&	X	&	X	&	X	\\
541.40717	&	X	&	X	&	X	&	X	&	X	&	X	\\
542.52497	&	X	&	X	&	X	&	X	&	N	&	X	\\
549.82172	&	X	&	X	&	X	&	X	&	X	&	X	\\
551.98345	&	X	&	X	&	O	&	X	&	O	&	X	\\
552.51168	&	X	&	X	&	X	&	X	&	X	&	X	\\
553.48393	&	X	&	X	&	X	&	X	&	X	&	X	\\
554.52601	&	X	&	X	&	O	&	X	&	O	&	X	\\
559.13604	&	X	&	X	&	X	&	X	&	X	&	X	\\
562.74892	&	X	&	X	&	O	&	X	&	O	&	O	\\
569.61110	&	X	&	X	&	X	&	X	&	N	&	X	\\
572.59548	&	X	&	X	&	X	&	X	&	X	&	X	\\
573.27177	&	X	&	X	&	X	&	X	&	X	&	X	\\
573.76905	&	X	&	X	&	X	&	X	&	X	&	X	\\
581.36693	&	X	&	X	&	O	&	X	&	N	&	N	\\
582.31562	&	X	&	X	&	O	&	X	&	N	&	N	\\
582.44054	&	X	&	X	&	O	&	X	&	N	&	N	\\
582.61222	&	X	&	X	&	O	&	X	&	N	&	N	\\
595.25188	&	X	&	X	&	O	&	X	&	N	&	X	\\
599.13724	&	X	&	X	&	X	&	X	&	N	&	X	\\
608.41036	&	X	&	X	&	X	&	X	&	N	&	X	\\
611.33234	&	X	&	X	&	O	&	X	&	N	&	X	\\
614.92483	&	X	&	X	&	O	&	X	&	N	&	X	\\
617.93936	&	X	&	X	&	X	&	X	&	N	&	X	\\
623.35389	&	X	&	X	&	X	&	X	&	N	&	X	\\
623.83872	&	O	&	O	&	X	&	X	&	N	&	X	\\
623.99439	&	X	&	X	&	X	&	X	&	N	&	X	\\
624.75582	&	(X)	&	(X)	&	(X)	&	(X)	&	N	&	O	\\
624.75769	&	(X)	&	(X)	&	(X)	&	(X)	&	N	&	O	\\
624.89067	&	X	&	X	&	X	&	X	&	N	&	X	\\
633.19566	&	X	&	X	&	X	&	X	&	N	&	X	\\
636.94579	&	X	&	X	&	X	&	X	&	N	&	X	\\
638.37294	&	X	&	X	&	X	&	X	&	N	&	X	\\
638.54591	&	X	&	X	&	X	&	X	&	N	&	X	\\
640.72458	&	X	&	X	&	X	&	X	&	N	&	X	\\
641.69204	&	X	&	X	&	X	&	X	&	N	&	X	\\
643.26762	&	X	&	X	&	X	&	X	&	N	&	X	\\
644.29559	&	X	&	X	&	X	&	X	&	N	&	X	\\
644.64059	&	X	&	X	&	X	&	X	&	N	&	X	\\
645.63809	&	X	&	X	&	X	&	X	&	N	&	X	\\
648.22074	&	X	&	X	&	O	&	X	&	N	&	X	\\
649.12537	&	X	&	X	&	O	&	X	&	N	&	X	\\
650.63403	&	X	&	X	&	X	&	X	&	N	&	X	\\
651.60767	&	X	&	X	&	X	&	X	&	N	&	X	\\
722.23923	&	X	&	X	&	O	&	N	&	N	&	X	\\
722.44790	&	X	&	X	&	O	&	N	&	N	&	X	\\
730.15647	&	X	&	X	&	X	&	N	&	N	&	X	\\
744.93305	&	X	&	O	&	X	&	N	&	N	&	X	\\
747.97024	&	X	&	X	&	X	&	N	&	N	&	X	\\
751.58309	&	X	&	X	&	X	&	N	&	N	&	X	\\
771.17205	&	X	&	X	&	O	&	N	&	N	&	X	\\
784.13864	&	X	&	X	&	X	&	N	&	N	&	X	\\

\end{longtable}
}
% End onllongtab

\onllongtab{3}{
\begin{longtable}{ccc}
\caption{\ion{Ti}{i} spectral lines examined and selected for the plots. X denotes accepted, and O rejected lines.  Laboratory wavelengths in air are from Forsberg (1991), with one additional decimal extracted from the original data by Litz\'{e}n (private comm.)}\\
\hline
\hline
$\lambda$ [nm] & Solar disk center & \object{Procyon} UVES 
\\
\hline
\endfirsthead
\caption{Continued.} \\
\hline
$\lambda$ [nm] & Solar disk center & Procyon UVES
\\
\hline
\endhead
\hline
\endfoot
\hline
\endlastfoot

400.80533	&	O	&	O	\\
400.89273	&	X	&	O	\\
401.32412	&	O	&	X	\\
401.53738	&	X	&	O	\\
401.62761	&	X	&	O	\\
401.77674	&	X	&	O	\\
402.18151	&	O	&	X	\\
402.45709	&	X	&	X	\\
402.65374	&	X	&	O	\\
403.05111	&	O	&	X	\\
403.39019	&	X	&	O	\\
403.48905	&	O	&	O	\\
404.03166	&	O	&	O	\\
405.29214	&	X	&	X	\\
405.50112	&	O	&	X	\\
406.00753	&	X	&	O	\\
406.02621	&	X	&	X	\\
406.42097	&	O	&	O	\\
406.50952	&	O	&	X	\\
406.55859	&	O	&	O	\\
407.84705	&	X	&	O	\\
408.24551	&	O	&	X	\\
409.91650	&	X	&	O	\\
411.27083	&	X	&	X	\\
412.21427	&	X	&	X	\\
413.72842	&	O	&	O	\\
414.30423	&	X	&	O	\\
415.07991	&	O	&	O	\\
415.09587	&	O	&	X	\\
415.96365	&	X	&	X	\\
416.63085	&	O	&	X	\\
416.93260	&	X	&	X	\\
417.10168	&	O	&	O	\\
417.40720	&	O	&	O	\\
417.73795	&	O	&	O	\\
418.32869	&	O	&	O	\\
418.61180	&	X	&	X	\\
418.86831	&	O	&	X	\\
420.07448	&	O	&	O	\\
421.17204	&	O	&	O	\\
424.91124	&	X	&	O	\\
425.16022	&	O	&	O	\\
425.17563	&	O	&	X	\\
425.85248	&	O	&	O	\\
426.07460	&	O	&	O	\\
426.31335	&	X	&	O	\\
426.52592	&	X	&	X	\\
426.57027	&	O	&	X	\\
426.62092	&	X	&	X	\\
426.89159	&	X	&	O	\\
427.01267	&	O	&	X	\\
427.32950	&	O	&	O	\\
427.43825	&	X	&	O	\\
427.45882	&	X	&	O	\\
427.64269	&	X	&	X	\\
427.66318	&	O	&	X	\\
427.82204	&	O	&	X	\\
428.00532	&	O	&	O	\\
428.13641	&	X	&	X	\\
428.26990	&	X	&	O	\\
428.49811	&	O	&	X	\\
428.60039	&	O	&	X	\\
428.74027	&	X	&	X	\\
428.81795	&	O	&	O	\\
428.90703	&	X	&	O	\\
429.09291	&	X	&	O	\\
429.26685	&	X	&	X	\\
429.57489	&	X	&	O	\\
429.86656	&	X	&	O	\\
429.92030	&	O	&	O	\\
429.96293	&	O	&	X	\\
430.05543	&	O	&	X	\\
430.10793	&	X	&	X	\\
430.59080	&	O	&	X	\\
431.47998	&	O	&	O	\\
431.86289	&	O	&	O	\\
432.16537	&	X	&	O	\\
432.63513	&	X	&	X	\\
433.48278	&	O	&	O	\\
433.84647	&	O	&	O	\\
434.61027	&	O	&	O	\\
436.04824	&	O	&	O	\\
439.39224	&	X	&	O	\\
439.48753	&	O	&	O	\\
440.42746	&	X	&	X	\\
440.56802	&	O	&	X	\\
441.24216	&	X	&	O	\\
441.65310	&	O	&	O	\\
441.72739	&	O	&	X	\\
442.17520	&	X	&	O	\\
442.28198	&	X	&	O	\\
442.70981	&	X	&	O	\\
443.00167	&	O	&	O	\\
443.03644	&	X	&	O	\\
443.65859	&	X	&	O	\\
443.66749	&	X	&	O	\\
444.03383	&	X	&	O	\\
444.12662	&	X	&	O	\\
444.91427	&	X	&	X	\\
445.08943	&	X	&	O	\\
445.33130	&	X	&	X	\\
445.36991	&	X	&	X	\\
445.53176	&	O	&	X	\\
445.74267	&	X	&	O	\\
446.33797	&	O	&	X	\\
446.35329	&	X	&	O	\\
446.58056	&	X	&	O	\\
447.12366	&	X	&	X	\\
447.48516	&	X	&	O	\\
448.05886	&	X	&	X	\\
448.12588	&	O	&	O	\\
448.26869	&	O	&	O	\\
448.50759	&	X	&	O	\\
449.25315	&	X	&	O	\\
449.61448	&	O	&	X	\\
450.37566	&	X	&	X	\\
451.27335	&	O	&	X	\\
451.37141	&	X	&	O	\\
451.80224	&	X	&	X	\\
451.86886	&	X	&	O	\\
452.27966	&	X	&	O	\\
452.73050	&	O	&	O	\\
452.74692	&	O	&	O	\\
453.32396	&	X	&	X	\\
453.47759	&	X	&	X	\\
453.55682	&	X	&	O	\\
453.59183	&	X	&	O	\\
453.60399	&	X	&	O	\\
454.46873	&	X	&	X	\\
454.77941	&	O	&	X	\\
454.87631	&	X	&	X	\\
455.24535	&	X	&	X	\\
455.54839	&	O	&	X	\\
455.81095	&	X	&	X	\\
455.99175	&	X	&	O	\\
456.26263	&	X	&	X	\\
456.34230	&	X	&	X	\\
461.72682	&	X	&	X	\\
461.95198	&	X	&	X	\\
462.30972	&	O	&	X	\\
462.93360	&	X	&	X	\\
463.93606	&	X	&	X	\\
463.96610	&	X	&	X	\\
463.99401	&	X	&	X	\\
464.51890	&	X	&	X	\\
465.00099	&	X	&	X	\\
465.56864	&	O	&	O	\\
465.60427	&	X	&	O	\\
465.64680	&	O	&	X	\\
466.75843	&	X	&	O	\\
467.51165	&	X	&	X	\\
468.19083	&	X	&	X	\\
468.68595	&	X	&	O	\\
469.07947	&	X	&	O	\\
469.87600	&	X	&	O	\\
470.89523	&	O	&	O	\\
471.52995	&	X	&	O	\\
472.26057	&	X	&	X	\\
473.11646	&	X	&	O	\\
473.34211	&	X	&	O	\\
474.21077	&	X	&	O	\\
474.27887	&	X	&	X	\\
474.76732	&	X	&	O	\\
475.81183	&	X	&	X	\\
475.88997	&	O	&	O	\\
475.92690	&	X	&	X	\\
476.97666	&	O	&	O	\\
477.10964	&	O	&	X	\\
477.82549	&	X	&	X	\\
478.17110	&	X	&	X	\\
479.24835	&	O	&	O	\\
479.62113	&	O	&	X	\\
479.79747	&	O	&	O	\\
479.97950	&	X	&	O	\\
480.19085	&	X	&	X	\\
480.54151	&	O	&	O	\\
480.85258	&	X	&	O	\\
481.10655	&	O	&	X	\\
481.28918	&	X	&	O	\\
481.90285	&	X	&	O	\\
482.04104	&	X	&	X	\\
482.54392	&	O	&	O	\\
482.75725	&	O	&	X	\\
483.20678	&	O	&	O	\\
483.73962	&	X	&	O	\\
484.08732	&	X	&	X	\\
484.39753	&	O	&	X	\\
484.84767	&	X	&	O	\\
485.60100	&	X	&	O	\\
486.82593	&	X	&	X	\\
487.01270	&	X	&	X	\\
488.09075	&	O	&	X	\\
488.50811	&	X	&	X	\\
489.30522	&	X	&	O	\\
489.34086	&	X	&	O	\\
489.99086	&	X	&	O	\\
490.05838	&	O	&	O	\\
490.09970	&	O	&	O	\\
490.37369	&	X	&	O	\\
490.90975	&	X	&	O	\\
491.36154	&	X	&	X	\\
491.52299	&	X	&	X	\\
491.98609	&	X	&	X	\\
492.17652	&	X	&	X	\\
492.54001	&	X	&	O	\\
492.61510	&	X	&	X	\\
492.83369	&	X	&	X	\\
493.77266	&	O	&	O	\\
494.15685	&	X	&	O	\\
494.30663	&	X	&	O	\\
494.81883	&	X	&	O	\\
495.82774	&	O	&	X	\\
496.47157	&	X	&	O	\\
497.53416	&	X	&	X	\\
497.81868	&	X	&	X	\\
498.17308	&	X	&	X	\\
498.91336	&	X	&	O	\\
499.10672	&	X	&	O	\\
499.50658	&	X	&	O	\\
499.70930	&	X	&	X	\\
499.95015	&	X	&	X	\\
500.09883	&	X	&	X	\\
500.72061	&	X	&	X	\\
500.96447	&	X	&	X	\\
501.32817	&	O	&	X	\\
501.41859	&	X	&	O	\\
501.61609	&	X	&	X	\\
502.00246	&	O	&	X	\\
502.28664	&	X	&	X	\\
502.48437	&	X	&	X	\\
502.55696	&	O	&	X	\\
503.59027	&	X	&	X	\\
503.64637	&	X	&	X	\\
503.83966	&	X	&	O	\\
503.99554	&	X	&	O	\\
504.06111	&	X	&	O	\\
504.35818	&	X	&	X	\\
504.54145	&	X	&	X	\\
504.82087	&	O	&	O	\\
505.28693	&	X	&	X	\\
505.40750	&	X	&	X	\\
506.21006	&	X	&	X	\\
506.40558	&	X	&	X	\\
506.46515	&	X	&	X	\\
506.59879	&	X	&	X	\\
506.83236	&	O	&	X	\\
507.14668	&	O	&	X	\\
508.53327	&	X	&	O	\\
508.70603	&	X	&	X	\\
510.94305	&	X	&	O	\\
511.34396	&	X	&	X	\\
512.04154	&	X	&	O	\\
513.29294	&	X	&	O	\\
514.54596	&	X	&	X	\\
514.74766	&	X	&	X	\\
515.21841	&	X	&	X	\\
517.37404	&	X	&	X	\\
518.63276	&	X	&	X	\\
518.95782	&	X	&	X	\\
519.29689	&	X	&	X	\\
519.40341	&	X	&	X	\\
520.10812	&	X	&	X	\\
521.03840	&	X	&	X	\\
521.12151	&	O	&	X	\\
521.29928	&	X	&	X	\\
521.96994	&	X	&	X	\\
522.26730	&	X	&	X	\\
522.36205	&	X	&	X	\\
522.43046	&	X	&	X	\\
522.45438	&	X	&	O	\\
522.49329	&	X	&	X	\\
523.09721	&	X	&	X	\\
523.85743	&	O	&	X	\\
524.61266	&	X	&	X	\\
524.65460	&	X	&	O	\\
524.72881	&	X	&	O	\\
524.83856	&	O	&	X	\\
525.09237	&	X	&	O	\\
525.14781	&	X	&	O	\\
525.20986	&	X	&	O	\\
525.99734	&	X	&	X	\\
526.59631	&	O	&	X	\\
527.16095	&	X	&	X	\\
528.23745	&	O	&	X	\\
528.43740	&	O	&	X	\\
528.87941	&	X	&	O	\\
528.92667	&	X	&	X	\\
529.57747	&	X	&	X	\\
530.00105	&	O	&	X	\\
531.32361	&	X	&	X	\\
533.83052	&	O	&	X	\\
534.06625	&	X	&	X	\\
535.10672	&	X	&	X	\\
536.66373	&	X	&	X	\\
538.46291	&	X	&	X	\\
538.91640	&	X	&	O	\\
538.99883	&	X	&	O	\\
539.10576	&	X	&	O	\\
539.65963	&	O	&	X	\\
539.70624	&	X	&	X	\\
542.62480	&	X	&	X	\\
542.91403	&	X	&	X	\\
543.23338	&	X	&	O	\\
543.83061	&	X	&	X	\\
544.66188	&	O	&	O	\\
544.79118	&	O	&	O	\\
544.89110	&	X	&	X	\\
544.91509	&	X	&	O	\\
545.36387	&	X	&	X	\\
546.04983	&	X	&	O	\\
547.11968	&	O	&	X	\\
547.26866	&	X	&	X	\\
547.35439	&	X	&	O	\\
547.42233	&	X	&	X	\\
547.44511	&	X	&	X	\\
547.76934	&	X	&	X	\\
548.14092	&	O	&	O	\\
548.18595	&	X	&	X	\\
548.81934	&	O	&	X	\\
549.01480	&	X	&	X	\\
549.47199	&	O	&	O	\\
550.38974	&	X	&	X	\\
551.17784	&	X	&	X	\\
551.25239	&	X	&	O	\\
551.43434	&	X	&	O	\\
551.45320	&	X	&	O	\\
551.80580	&	X	&	O	\\
560.07851	&	O	&	O	\\
564.41326	&	X	&	O	\\
564.85619	&	X	&	X	\\
565.90951	&	X	&	O	\\
566.21484	&	X	&	X	\\
566.28812	&	O	&	X	\\
567.30465	&	X	&	X	\\
567.34187	&	X	&	O	\\
567.54085	&	X	&	X	\\
567.99141	&	X	&	X	\\
568.94590	&	O	&	X	\\
570.26585	&	X	&	X	\\
571.38805	&	X	&	X	\\
571.51143	&	O	&	X	\\
571.64420	&	X	&	X	\\
572.04353	&	X	&	X	\\
573.90492	&	X	&	X	\\
573.94683	&	X	&	X	\\
573.99778	&	X	&	X	\\
574.11971	&	X	&	X	\\
574.44618	&	X	&	X	\\
576.63262	&	X	&	O	\\
577.40259	&	X	&	O	\\
577.44751	&	O	&	O	\\
578.07599	&	O	&	O	\\
578.59768	&	X	&	O	\\
578.97483	&	X	&	O	\\
579.74299	&	X	&	O	\\
580.42625	&	X	&	O	\\
581.28198	&	X	&	O	\\
582.36864	&	X	&	O	\\
583.24733	&	X	&	O	\\
584.11732	&	X	&	O	\\
586.64492	&	X	&	X	\\
587.76573	&	O	&	O	\\
588.02679	&	X	&	O	\\
589.92915	&	X	&	O	\\
590.33146	&	X	&	O	\\
590.64204	&	O	&	O	\\
591.85338	&	X	&	X	\\
592.21075	&	X	&	O	\\
593.78069	&	X	&	O	\\
594.06459	&	X	&	O	\\
594.17500	&	X	&	X	\\
595.31572	&	X	&	X	\\
596.58246	&	X	&	O	\\
597.85385	&	X	&	X	\\
598.08222	&	O	&	X	\\
598.25298	&	X	&	O	\\
598.85515	&	X	&	O	\\
601.75453	&	X	&	O	\\
601.86361	&	X	&	O	\\
606.46234	&	X	&	O	\\
608.52246	&	X	&	X	\\
609.11695	&	X	&	X	\\
609.27899	&	X	&	O	\\
609.86573	&	X	&	X	\\
612.09989	&	X	&	O	\\
612.62142	&	O	&	X	\\
614.62072	&	X	&	O	\\
614.97245	&	X	&	O	\\
617.47513	&	X	&	O	\\
622.04721	&	X	&	O	\\
622.14302	&	O	&	O	\\
625.80992	&	X	&	X	\\
625.87053	&	X	&	X	\\
626.10960	&	X	&	X	\\
626.60093	&	X	&	O	\\
626.85222	&	X	&	X	\\
630.37538	&	X	&	X	\\
631.12808	&	O	&	X	\\
631.22344	&	X	&	X	\\
631.80443	&	O	&	X	\\
632.51621	&	O	&	X	\\
633.60961	&	X	&	X	\\
641.31016	&	X	&	O	\\
641.90873	&	X	&	X	\\
650.81188	&	O	&	O	\\
654.62656	&	O	&	X	\\
655.42222	&	X	&	X	\\
655.60605	&	X	&	X	\\
659.29143	&	X	&	X	\\
659.91049	&	X	&	X	\\
667.85687	&	X	&	X	\\
671.66655	&	X	&	X	\\
674.31195	&	X	&	X	\\
674.55459	&	X	&	X	\\
674.63298	&	X	&	O	\\
699.66366	&	X	&	X	\\
706.90730	&	X	&	O	\\
713.80737	&	X	&	X	\\
713.89025	&	X	&	X	\\
718.98767	&	O	&	O	\\
721.61822	&	X	&	X	\\
725.17079	&	X	&	X	\\
735.21195	&	X	&	O	\\
735.77258	&	X	&	X	\\
736.40990	&	X	&	O	\\
744.05761	&	X	&	O	\\
749.61032	&	X	&	X	\\
758.02650	&	X	&	O	\\
794.91506	&	X	&	O	\\

\end{longtable}
}

\onllongtab{4}{
\begin{longtable}{ccc}
\caption{\ion{Ti}{ii} spectral lines examined and selected for the plots. X denotes accepted, and O rejected lines.  Laboratory wavelengths in air are from Zapadlik et al. (1995), Zapadlik (1996), and Johansson (private comm.) }\\
\hline
\hline
$\lambda$ [nm] & Solar disk center & Integrated sunlight 
\\
\hline
\endfirsthead
\caption{Continued.} \\
\hline
$\lambda$ [nm] & Solar disk center & Integrated sunlight
\\
\hline
\endhead
\hline
\endfoot
\hline
\endlastfoot

401.23833	&	X	&	O	\\
402.51295	&	X	&	X	\\
402.83397	&	X	&	X	\\
405.38225	&	O	&	O	\\
416.15300	&	O	&	O	\\
416.36435	&	X	&	X	\\
417.19032	&	X	&	X	\\
428.78749	&	O	&	O	\\
429.02143	&	X	&	X	\\
429.40934	&	O	&	O	\\
430.00414	&	X	&	X	\\
430.19225	&	O	&	O	\\
430.78646	&	O	&	O	\\
431.28590	&	O	&	O	\\
431.49698	&	O	&	O	\\
431.67930	&	X	&	X	\\
432.09505	&	X	&	X	\\
433.02337	&	X	&	X	\\
433.06973	&	X	&	X	\\
433.79141	&	X	&	X	\\
434.13554	&	O	&	O	\\
434.42792	&	X	&	X	\\
435.08342	&	O	&	O	\\
437.48124	&	X	&	X	\\
438.68445	&	X	&	X	\\
439.40586	&	X	&	X	\\
439.50310	&	X	&	X	\\
439.58375	&	X	&	X	\\
439.97644	&	O	&	O	\\
441.10730	&	X	&	X	\\
441.77128	&	X	&	O	\\
441.83295	&	X	&	O	\\
444.17297	&	O	&	O	\\
444.37997	&	X	&	X	\\
444.45568	&	X	&	X	\\
445.04824	&	X	&	X	\\
446.44482	&	X	&	X	\\
446.84916	&	X	&	X	\\
448.83251	&	X	&	X	\\
450.12691	&	X	&	X	\\
452.94791	&	O	&	O	\\
453.39596	&	X	&	X	\\
454.40112	&	X	&	O	\\
454.96210	&	X	&	X	\\
454.98117	&	X	&	X	\\
456.37554	&	X	&	O	\\
457.19703	&	X	&	X	\\
458.99561	&	O	&	O	\\
465.72018	&	X	&	X	\\
470.86627	&	X	&	X	\\
477.99801	&	X	&	X	\\
479.85332	&	X	&	X	\\
480.50910	&	X	&	O	\\
486.56091	&	O	&	O	\\
487.40096	&	X	&	X	\\
491.11947	&	X	&	X	\\
501.02091	&	O	&	O	\\
506.90909	&	X	&	X	\\
507.22843	&	X	&	X	\\
512.91484	&	X	&	X	\\
515.40729	&	X	&	X	\\
518.58993	&	X	&	O	\\
518.86848	&	X	&	X	\\
521.15378	&	O	&	O	\\
522.65374	&	X	&	X	\\
526.86048	&	X	&	O	\\
533.67851	&	X	&	X	\\
538.10218	&	X	&	X	\\
541.87632	&	X	&	X	\\
649.15652	&	O	&	O	\\
655.95673	&	X	&	X	\\
660.69460	&	X	&	X	\\
721.47288	&	O	&	O	\\

\end{longtable}
}

\onllongtab{5}{
\begin{longtable}{ccc}
\caption{\ion{Cr}{ii} spectral lines examined and selected for the plots. X denotes accepted, and O rejected lines.  Laboratory wavelengths in air are from the ongoing FERRUM Project (Johansson 2002, \& private comm.) }\\
\hline
\hline
$\lambda$ [nm] & Solar disk center & Integrated sunlight 
\\
\hline
\endfirsthead
\caption{Continued.} \\
\hline
$\lambda$ [nm] & Solar disk center & Integrated sunlight
\\
\hline
\endhead
\hline
\endfoot
\hline
\endlastfoot

404.91000	&	O	&	O	\\
405.19357	&	O	&	O	\\
405.40793	&	X	&	O	\\
408.61290	&	X	&	X	\\
408.75934	&	X	&	X	\\
408.88332	&	X	&	X	\\
411.09851	&	O	&	O	\\
411.10091	&	O	&	O	\\
411.25465	&	X	&	X	\\
411.32158	&	O	&	O	\\
413.24144	&	O	&	O	\\
414.57764	&	O	&	O	\\
417.06178	&	X	&	X	\\
417.94212	&	O	&	O	\\
422.48538	&	X	&	X	\\
424.23660	&	O	&	O	\\
424.63997	&	X	&	X	\\
425.26255	&	O	&	O	\\
426.92678	&	O	&	O	\\
427.55610	&	O	&	O	\\
428.41854	&	X	&	X	\\
455.49892	&	O	&	O	\\
455.86440	&	X	&	X	\\
458.81984	&	X	&	X	\\
459.20525	&	X	&	X	\\
461.66240	&	X	&	X	\\
461.88067	&	O	&	O	\\
463.40727	&	X	&	X	\\
481.23441	&	X	&	O	\\
482.41306	&	O	&	O	\\
483.62294	&	X	&	X	\\
484.82494	&	X	&	X	\\
485.61850	&	X	&	X	\\
486.02127	&	O	&	O	\\
486.43299	&	O	&	O	\\
487.63920	&	O	&	O	\\
488.46039	&	X	&	X	\\
509.73208	&	X	&	X	\\
521.08663	&	O	&	O	\\
523.24980	&	X	&	O	\\
523.73214	&	X	&	X	\\
524.67740	&	X	&	X	\\
524.94406	&	X	&	X	\\
527.49742	&	O	&	O	\\
527.98763	&	X	&	X	\\
528.00691	&	X	&	X	\\
530.58642	&	X	&	X	\\
530.84210	&	X	&	X	\\
531.06922	&	X	&	X	\\
531.35808	&	X	&	X	\\
533.48679	&	X	&	X	\\
533.77829	&	O	&	O	\\
534.60860	&	X	&	X	\\
534.65463	&	O	&	O	\\
536.93502	&	X	&	X	\\
540.76149	&	X	&	X	\\
542.09300	&	X	&	X	\\

\end{longtable}
}

\onllongtab{6}{
\begin{longtable}{cccc}
\caption{\ion{Ca}{i} spectral lines examined and selected for the plots. X denotes accepted, and O rejected lines.  Laboratory wavelengths in air are from Litz\'{e}n (private comm.) }\\
\hline
\hline
$\lambda$ [nm] & Solar disk center & Integrated sunlight & \object{Procyon} UVES 
\\
\hline
\endfirsthead
\caption{Continued.} \\
\hline
$\lambda$ [nm] & Solar disk center & Integrated sunlight & Procyon UVES 
\\
\hline
\endhead
\hline
\endfoot
\hline
\endlastfoot

422.67270	&	O	&	O	&	O	\\
428.30094	&	O	&	O	&	X	\\
428.93634	&	O	&	O	&	X	\\
429.89863	&	O	&	O	&	O	\\
430.25263	&	O	&	O	&	O	\\
430.77409	&	O	&	O	&	O	\\
431.86495	&	X	&	O	&	O	\\
435.50807	&	X	&	O	&	X	\\
442.54353	&	O	&	O	&	O	\\
443.49566	&	X	&	O	&	O	\\
443.56791	&	O	&	O	&	O	\\
445.58863	&	O	&	O	&	X	\\
445.66144	&	O	&	O	&	X	\\
452.69281	&	O	&	O	&	X	\\
457.85523	&	X	&	X	&	X	\\
458.13959	&	O	&	O	&	O	\\
458.58646	&	X	&	O	&	O	\\
534.94615	&	X	&	X	&	X	\\
551.29808	&	X	&	O	&	X	\\
558.19681	&	X	&	X	&	X	\\
558.87511	&	O	&	O	&	X	\\
559.01132	&	X	&	X	&	X	\\
559.44611	&	X	&	O	&	O	\\
559.84798	&	O	&	O	&	O	\\
560.12746	&	O	&	O	&	X	\\
560.28407	&	O	&	O	&	O	\\
585.74495	&	X	&	X	&	X	\\
586.75607	&	X	&	O	&	X	\\
610.27178	&	X	&	O	&	X	\\
612.22148	&	X	&	X	&	X	\\
616.12897	&	O	&	O	&	X	\\
616.21704	&	X	&	X	&	X	\\
616.37498	&	X	&	X	&	X	\\
616.64334	&	X	&	X	&	X	\\
616.90342	&	X	&	X	&	X	\\
616.95548	&	X	&	O	&	X	\\
643.90718	&	X	&	X	&	X	\\
644.98089	&	X	&	X	&	X	\\
645.55974	&	X	&	X	&	X	\\
646.25643	&	X	&	O	&	O	\\
647.16581	&	X	&	X	&	X	\\
649.37783	&	X	&	X	&	X	\\
649.96463	&	X	&	X	&	X	\\
657.27789	&	X	&	X	&	X	\\
671.76773	&	O	&	O	&	X	\\
714.81455	&	X	&	X	&	X	\\
720.21914	&	X	&	X	&	X	\\

\end{longtable}
}


\begin{thebibliography}{}

  \bibitem[1999]{allende99} Allende Prieto, C.\ 1999,
  The McDonald Observatory Spectrum of Procyon between 4559 and 5780 \AA, \texttt{http://hebe.as.utexas.edu/procyon}   (`Procyon McDonald')

   \bibitem[1998]{allendeetal98} Allende Prieto, C., \& Garc{\'\i}a L\'{o}pez, R.\ 1998, A\&AS 131, 431
   
   \bibitem[1999]{allendeetal99} Allende Prieto, C., Garc{\'\i}a L\'{o}pez, R., Lambert, D.\ L., \& Gustafsson, B.\ 1999, ApJ 526, 991

  \bibitem[2002a]{allendeetal02a} Allende Prieto, C., Asplund, M., Garc{\'\i}a L\'{o}pez, R.\ J., \& Lambert, D. L.\ 2002a, ApJ 567, 544
  
   \bibitem[2002b]{allendeetal02b} Allende Prieto, C., Lambert, D.\ L., Tull, R.\ G., \& MacQueen, P.\ J.\ 2002b, ApJ 566, L93

   \bibitem[2005]{andersen05} Andersen, M.\ I., Strassmeier, K.\ G., Hoffman, A., Woche, M., \& Spano, P.\ 2005, in High Resolution Infrared Spectroscopy in Astronomy, ed. H.~U.~K\"{a}ufl, R.~Siebenmorgen, \& A.~F.~M.~Moorwood (Springer, Berlin), 57

   \bibitem[2007]{araujo07} Araujo-Hauck, C., Pasquini, L., Manescau, A., \& et al.\ 2007, ESO Messenger 129, 24
   
   \bibitem[2005]{asplund05a} Asplund, M.\ 2005, ARA\&A 43, 481
   
   \bibitem[2000]{asplund00a} Asplund, M., Ludwig, H.-G., Nordlund, \AA., \& Stein, R.\ F.\ 2000a, A\&A 359, 669

   \bibitem[2000]{asplund00b} Asplund, M., Nordlund, \AA., Trampedach, R., Allende Prieto, C., \& Stein, R.\ F.\ 2000b, A\&A 359, 729
   
   \bibitem[2005]{asplund05b} Asplund, M., Grevesse, N., Sauval, A.\ J., Allende Prieto, C., \& Blomme, R.\ 2005, A\&A 431, 693
   
   \bibitem[2006]{asplund06} Asplund, M., Lambert, D.\ L., Nissen, P.\ E., Primas, F., \& Smith, V.\ V.\ 2006, ApJ 644, 229
   
   \bibitem[1989]{atrosh89} Atroshchenko, I.\ N., Gadun, A.\ S., \& Kostyk, R.\ I.\ 1989, Astrofizika 31, 589 =  Astrophysics 31, 765 (1990)

   \bibitem[2005]{aufden05} Aufdenberg, J.\ P., Ludwig, H.-G., \& Kervella, P.\ 2005, ApJ 633, 424
   
   \bibitem[1975]{ayres75} Ayres, T.\ R., \& Linsky, J.\ L.\ 1975, ApJ 201, 212
   
   \bibitem[2003]{bagnulo03} Bagnulo, S., Jehin, E., Ledoux, C., \& et al.\ 2003, ESO Messenger 114, 10,\\ \texttt{http://www.sc.eso.org/santiago/uvespop} 
   
   \bibitem[2007]{bergemann07} Bergemann, M., \& Gehren, T.\ 2007, A\&A 473, 291
   
   \bibitem[2006]{bigot06} Bigot, L., Kervella, P., Th\'{e}venin, F., \& S\'{e}gransan, D.\ 2006, A\&A 446,635
   
   \bibitem[1985]{caccin85} Caccin, B., Cavallini, F., Ceppatelli, G., Righini, A., \& Sambuco, A.\ M.\ 1985, A\&A 149, 357
   
   \bibitem[2007]{cayrel07} Cayrel, R., Steffen, M., Chand, H., \& et al.\ 2007, A\&A 473, L37
   
   \bibitem[2007]{cheung07} Cheung, M.\ C.\ M., Sch\"{u}ssler, M., \& Moreno-Insertis, F.\ 2007, A\&A 461, 1163
     
   \bibitem[2000]{dekker00} Dekker, H., D'Odorico, S., Kaufer, A., Delabre, B., \& Kotzlowski, H.\ 2000, in Proc.\ SPIE 4008, Optical and IR Telescope Instrumentation and Detectors, ed. M.~Iye, \& A.~F.~Moorwood, 534
   
   \bibitem[1989]{delbouille89} Delbouille, L., Roland, G., \& Neven, L.\ 1989, Atlas photom\'{e}trique du spectre solaire de $\lambda$~3000 a $\lambda$~10000 (Li\`{e}ge: Universit\'{e} de Li\`{e}ge, Institut d'Astrophysique) [`Jungfraujoch atlas'] Digital version: BASS2000, BAse de donn\'{e}es Solaire Sol, Observatoire de Paris, \texttt{http://bass2000.obspm.fr}

   \bibitem[1995]{diego95} Diego, F., Fish, A.\ C., Barlow, M.\ J., \& et al.\ 1995, MNRAS 272, 323
   
   \bibitem[2007]{dodorico07} D'Odorico, V., \& the CODEX/ESPRESSO team 2007, Mem.Soc.Astron.Ital. 78, 712
   
   \bibitem[1982]{dravins82} Dravins, D.\ 1982, ARA\&A 20, 61
    
   \bibitem[1987]{dravins87a} Dravins, D.\ 1987a, A\&A 172, 200
   
   \bibitem[1987]{dravins87b} Dravins, D.\ 1987b, A\&A 172, 211
   
   \bibitem[1990] {dravins90} Dravins, D.\ 1990, A\&A 228, 218
   
   \bibitem[1994]{dravins94} Dravins, D.\ 1994, in NATO ASI Ser.~C, 436, The Impact of Long-Term Monitoring on Variable Star Research, ed. C.~Sterken, \& M.~De Groot (Kluwer, Dordrecht), 269 

   \bibitem[1990a]{dravins90a} Dravins, D., \& Nordlund, {\AA}.\ 1990a, A\&A 228, 184

   \bibitem[1990b]{dravins90b} Dravins, D., \& Nordlund, {\AA}.\ 1990b, A\&A 228, 203

   \bibitem[1981]{dravins81} Dravins, D., Lindegren, L., \& Nordlund, {\AA}.\ 1981, A\&A 96, 345
   
   \bibitem[1986]{dravins86} Dravins, D., Larsson, B., \& Nordlund, {\AA}.\ 1986, A\&A 158, 83
   
   \bibitem[1990]{dravins90c} Dravins, D., Lindegren, L., \& Torkelsson, U., {\AA}.\ 1990, A\&A 237, 137
   
   \bibitem[1999]{dravins99} Dravins, D., Lindegren, L., \& Madsen, S.\ 1999, A\&A 348, 1040
   
   \bibitem[2005]{dravins05} Dravins, D., Lindegren, L., Ludwig, H.-G., \& Madsen, S.\ 2005, in ESA SP-560, 13th Cambridge Workshop on Cool Stars, Stellar Systems and the Sun, ed. F.~Favata, G.~Hussain, \& B.~Battrick, 113
      
   \bibitem[1993]{edvardsson93} Edvardsson, B., Andersen, J., Gustafsson, B., \& et al.\ 1993, A\&A 275, 101
     
   \bibitem[1991]{forsberg91} Forsberg, P.\ 1991, Phys.Scr.\ 44, 446

   \bibitem[2005]{frutiger05} Frutiger, C., Solanki, S.\ K., \& Mathys, G.\ 2005, A\&A 444, 549
   
   \bibitem[1994]{gadun94} Gadun, A.\ S.\ 1994, Kin.Fiz.Nebesn.Tel 10 (3), 33 = Kin.Phys.Celestial Bodies 10 (3), 26
     
   \bibitem[2002]{ge02} Ge, J., Angel, J.\ R.\ P., Jacobsen,\ B., \& et al.\ 2002, PASP 114, 879

   \bibitem[2005]{gray05} Gray, D.\ F.\ 2005, PASP 117, 711

   \bibitem[2005]{griffin05} Griffin, R.\ E.\ 2005, PASP 117, 885
   
   \bibitem[1979]{griffin79} Griffin, R., \& Griffin, R.\ 1979, A photometric atlas of the spectrum of Procyon    3140-7470 {\AA} (Institute of Astronomy, Cambridge U.K.) [`Griffin Atlas']
  
   \bibitem[2006]{hadrava06} Hadrava, P.\ 2006, A\&A 448, 1149
  
   \bibitem[2000]{hinkle00} Hinkle, K., Wallace, L., Valenti, J., \& Harmer, D.\ 2000, Visible and Near Infrared Atlas of the Arcturus Spectrum 3727-9300~{\AA} (ASP, San Francisco) 

   \bibitem[2003]{hinkle03} Hinkle, K.\ H., Wallace, L., \& Livingston, W.\ 2003, Bull.AAS 35, 1260

   \bibitem[2003]{ivarsson03} Ivarsson, S., Wahlgren, G.\ M., \& Ludwig, H.-G.\ 2003, Bull.AAS 35, 1421
   
   \bibitem[2002]{johansson02} Johansson, S.\ 2002, IAU Highlights of Astronomy, 12, 84
    
   \bibitem[2005]{kurucz05} Kurucz, R.\ L.\ 2005, Mem.Soc.Astron.It.Suppl. 8, 189
   
   \bibitem[1984]{kurucz84} Kurucz, R.\ L., Furenlid, I., \& Brault, J.\ 1984, Solar flux atlas from 296 to 1300 nm (Sunspot, New Mexico: National Solar Observatory Atlas No.1, 1984) [`Kitt Peak Atlas'].  Digital version: National Solar Observatory Digital Library, \texttt{http://diglib.nso.edu} 
   
   \bibitem[1968]{lambert68} Lambert, D.\ L.\ 1968, MNRAS 138, 143
   
   \bibitem[2005]{leenaarts05} Leenaarts, J., \& Wedemeyer-B{\"{o}}hm, S.\ 2005, A\&A 431, 687

   \bibitem[2000]{lindegren00} Lindegren, L., Madsen, S., \& Dravins, D.\ 2000, A\&A 356, 1119
     
   \bibitem[1993]{litzen93} Litz\'{e}n, U., Brault, J.\ W., \& Thorne, A.\ P.\ 1993, Phys.Scr.\ 47, 628

   \bibitem[2007]{lovis07} Lovis, C., \& Pepe, F.\ 2007, A\&A 468, 1115
   
   \bibitem[2002]{madsen02} Madsen, S., Dravins, D. \& Lindegren, L.\ 2002, A\&A 381, 446

   \bibitem[2003]{madsen03} Madsen, S., Dravins, D., Ludwig, H.-G., \& Lindegren, L.\ 2003, A\&A 411, 581
   
   \bibitem[2008]{molaro08} Molaro, P., Levshakov, S.\ A., Monai, S., \& et al.\ 2008, A\&A 481, 559
   
   \bibitem[2007a]{murphy07a} Murphy, M.\ T., Tzanavaris, P., Webb, J.\ K., \& Lovis, C.\ 2007a, MNRAS 378, 221

   \bibitem[2007b]{murphy07b} Murphy, M.\ T., Udem, Th., Holzwarth, R., \& et al.\ 2007b, MNRAS 380, 839
   
   \bibitem[2007]{muthsam07} Muthsam, H.\ J., L\"{o}w-Baselli, B., Obertscheider, C., \& et al.\ 2007, MNRAS 380, 1335

   \bibitem[1994]{nave94} Nave, G., Johansson, S., Learner, R.\ C.\ M., Thorne, A.\ P., \& Brault, J.\ W.\ 1994, ApJS 94, 221

   \bibitem[1999]{neckel99} Neckel, H.\ 1999, Spectral Atlas of Solar Absolute Disk-Averaged and Disk-Center Intensity from 3290 to 12510 {\AA} (original version by J.~Brault \& H.~Neckel, 1987, unpublished) digital version from Hamburg Observatory; [`Hamburg photosphere']; also Solar Phys.\ 184, 421

   \bibitem[2000]{nissen00} Nissen, P.\ E.\ 2000, in IAU Symp.198, Metal Poor Stars, The Light Elements and their Evolution, ed. L.~da Silva, R.~de Medeiros, \& M.~Spite, 259 
  
   \bibitem[2005]{pasquini05} Pasquini, L., Cristiani, S., Dekker, H., \& et al.\ 2005, ESO Messenger 122, 10
   
   \bibitem[1996]{pickering96} Pickering, J.\ C., \& Thorne, A.\ P.\ 1996, ApJS 107, 761
   
   \bibitem[1973]{pierce73} Pierce, A.\ K., \& Breckinridge, J.\ B.\ 1973, Kitt Peak Nat.Obs.Contr. No. 559
   
   \bibitem[2005]{robinson05} Robinson, F.\ J., Demarque, P., Guenther, D. B., Kim, Y.-C., \& Chan, K.\ L.\ 2005, MNRAS 362, 1031
   
   \bibitem[2004]{rutten04} Rutten, R.\ J., de Wijn, A.\ G., \& S\"{u}tterlin, P.\ 2004, A\&A 416, 333

   \bibitem[1980]{rutten80} Rutten, R.\ J., \& Stencel, R.\ E.\ 1980, A\&AS 39, 415
   
   \bibitem[2004]{sacco04} Sacco, G.\ G., Pallavicini, R., Spano, P., \& et al.\ 2004, in Proc.\ SPIE 5490, Advancements in Adaptive Optics, ed. D.~B.~Calia, B.~L.~Ellerbroek, \& R.~Ragazzoni, 398

   \bibitem[2008]{schussler08} Sch\"{u}ssler, M., \& V\"{o}gler, A.\ 2008, A\&A 481, L5
    
   \bibitem[2005]{shchukina05} Shchukina, N.\ G., Trujillo Bueno, J., \& Asplund, M.\ 2005, ApJ 618, 939

   \bibitem[2005]{steffen05} Steffen, M., Freytag, B., \& Ludwig, H.-G.\ 2005, in ESA SP-560, 13th Cambridge Workshop on Cool Stars, Stellar Systems and the Sun, ed. F.~Favata, G.~Hussain, \& B.~Battrick, 985

   \bibitem[2007]{steffen07} Steffen, M., Strassmeier, K.\ G.\ 2007, Astron.Nachr.\ 328, 632
   
   \bibitem[2008]{steinmetz08} Steinmetz, T., Wilken, T., Araujo-Hauck, C., \& et al.\ 2008, Science 321, 1335
      
	 \bibitem[2003]{strassmeier03} Strassmeier, K.\ G., Hofmann, A., Woche, M.\ F., \& et al.\ 2003, in Proc.\ SPIE 4843, Polarimetry in Astronomy, ed. S.~Fineschi, 180
	 
   \bibitem[2004]{strassmeier04} Strassmeier, K.\ G., Pallavicini, R., Rice, J.\ B., \& Andersen, M.\ I.\ 2004, Astron.Nachr.\ 325, 278

   \bibitem[2007]{stubbs07} Stubbs, C.\ W., High, F.\ W., George, M.\ R., \& et al.\ 2007, PASP 119, 1163
   
   \bibitem[2006]{uitenbroek06} Uitenbroek, H.\ 2006, ApJ 639, 516

   \bibitem[2007]{vogler07} V\"{o}gler, A., \& Sch\"{u}ssler, M.\ 2007, A\&A 465, L43
     
   \bibitem[1986]{watanabe86} Watanabe, T., \& Steenbock, W.\ 1986, A\&A 165, 163
   
   \bibitem[1996]{wiese96} Wiese, W.\ L., Fuhr, J.\ R., \& Deters, T.\ M.\ 1996, J.Phys.Chem.Ref.Data, Monograph No.\ 7

   \bibitem[1996]{zapadlik96} Zapadlik, I.\ 1996, Studies of spectra recorded in space and in the laboratory, Licentiate thesis, Lund University

   \bibitem[1995]{zapadliketal95} Zapadlik, I., Johansson, S., Litz\'{e}n, U.\ 1995, in ASPC 81, Workshop on Laboratory and astronomical high resolution spectra, ed. A.~J.~Sauval, R.~Blomme, \& N.~Grevesse, 237 
 
\end{thebibliography}
\end{document}